\documentclass[
 superscriptaddress,
twocolumn,
 amsmath,amssymb,
 aps,
 prx,
 floatfix
]{revtex4-2}

\usepackage{booktabs}
\usepackage{graphicx}
\usepackage{dcolumn}
\usepackage{bm}
\usepackage{xcolor}
\usepackage{braket}
\usepackage{amsmath, amssymb,mathcomp,enumerate,amsfonts}
\usepackage{bbold}
\usepackage{soul}
\usepackage{siunitx}
\usepackage[colorlinks=true,allcolors=black]{hyperref}
\usepackage{comment}
\usepackage{array}
\usepackage[version=4]{mhchem}
\usepackage{mathtools}
\usepackage{float}

\DeclareSIUnit\bar{bar}

\usepackage{cleveref}
\crefname{equation}{Eq.}{Eqs.}
\Crefname{equation}{Equation}{Equations}
\crefname{figure}{Fig.}{Figs.}
\Crefname{figure}{Figure}{Figures}
\crefname{section}{Sec.}{Secs.}
\crefname{subsection}{Subsec.}{Subsecs.}
\Crefname{section}{Section}{Sections}
\crefname{appendix}{Appendix}{Apps.}
\Crefname{appendix}{Appendix}{Apps.}
\crefname{paragraph}{Sec.}{Secs.}
\crefname{table}{Table}{Tables}
\usepackage{bm}
\newcommand{\textalert}[1]{}

\newcommand{\eps}{\varepsilon}

\newcommand{\om}{\omega} 
\newcommand{\wdr}{\om_\text{d}} 

\newcommand{\half}{\frac{1}{2}}

\def\ie{i.e.\ }

\usepackage{xr}

\graphicspath{{nb_fig/}}

\usepackage{epstopdf}

\newcommand{\ECa}{E_{\text{C}a}}

\newcommand{\bna}{\hat{\bm{n}}_a}

\newcommand{\bnq}{\hat{\bm{n}}_q}

\newcommand{\bpha}{\hat{\bm{\varphi}}_a}

\newcommand{\bphq}{\hat{\bm{\varphi}}_q}
\newcommand{\phx}{\varphi_\text{ext}}

\newcommand{\ECq}{E_{Cq}}

\newcommand{\hH}{\hat{H}}

\newcommand{\wa}{\omega_a}

\newcommand{\wc}{\omega_c}

\newcommand{\uaa}{u_{aa}}

\newcommand{\uac}{u_{ac}}

\newcommand{\uca}{u_{ca}}

\newcommand{\ucc}{u_{cc}}

\newcommand{\Wa}{\bm{\omega}_a}

\newcommand{\Wc}{\bm{\omega}_c}
\newcommand{\Wq}{\bm{\omega}_q}
\newcommand{\ha}{\hat{a}}

\newcommand{\hc}{\hat{c}}

\newcommand{\hba}{\hat{\bm{a}}}

\newcommand{\hbc}{\hat{\bm{c}}}
\newcommand{\hbq}{\hat{\bm{q}}}
\newcommand{\ala}{\alpha_a}

\newcommand{\alc}{\alpha_c}

\newcommand{\bala}{\bm{\alpha}_{a}}
\newcommand{\balq}{\bm{\alpha}_{q}}

\begin{document}

\title{Suppression of measurement-induced state transitions in \texorpdfstring{cos$\varphi$}{cosϕ}-coupling transmon readout}

\author{Cyril Mori}
\affiliation{Univ. Grenoble Alpes, CNRS, Grenoble INP, Institut N\'eel, Grenoble, France}
\author{Francesca D'Esposito}
\affiliation{Univ. Grenoble Alpes, CNRS, Grenoble INP, Institut N\'eel, Grenoble, France}
\author{Alexandru Petrescu}
\affiliation{Laboratoire de Physique de l’Ecole Normale Supérieure, Mines Paris, Inria, CNRS, ENS-PSL, Sorbonne Université, PSL Research University, Paris, France}
\author{Lucas Ruela}
\affiliation{Univ. Grenoble Alpes, CNRS, Grenoble INP, Institut N\'eel, Grenoble, France}
\author{Shelender Kumar}
\affiliation{Univ. Grenoble Alpes, CNRS, Grenoble INP, Institut N\'eel, Grenoble, France}
\author{Vishnu Narayanan Suresh}
\affiliation{Univ. Grenoble Alpes, CNRS, Grenoble INP, Institut N\'eel, Grenoble, France}
\author{Waël Ardati}
\affiliation{Univ. Grenoble Alpes, CNRS, Grenoble INP, Institut N\'eel, Grenoble, France}
\author{Dorian Nicolas}
\affiliation{Univ. Grenoble Alpes, CNRS, Grenoble INP, Institut N\'eel, Grenoble, France}
\author{Giulio Cappelli}
\affiliation{Univ. Grenoble Alpes, CNRS, Grenoble INP, Institut N\'eel, Grenoble, France}
\author{Arpit Ranadive}
\altaffiliation{Present address: Google Quantum AI, Goleta, California 93117, USA.}
\affiliation{Univ. Grenoble Alpes, CNRS, Grenoble INP, Institut N\'eel, Grenoble, France}
\author{Gwenael Le Gal}
\affiliation{Univ. Grenoble Alpes, CNRS, Grenoble INP, Institut N\'eel, Grenoble, France}
\author{Martina Esposito}
\altaffiliation{Present address: CNR-SPIN Complesso di Monte S. Angelo, 80126 Napoli, Italy.}
\affiliation{Univ. Grenoble Alpes, CNRS, Grenoble INP, Institut N\'eel, Grenoble, France}
\author{Quentin Ficheux}
\affiliation{Univ. Grenoble Alpes, CNRS, Grenoble INP, Institut N\'eel, Grenoble, France}
\author{Nicolas Roch}
\affiliation{Univ. Grenoble Alpes, CNRS, Grenoble INP, Institut N\'eel, Grenoble, France}
\author{Olivier Buisson}
\email{Contact author: olivier.buisson@neel.cnrs.fr}
\affiliation{Univ. Grenoble Alpes, CNRS, Grenoble INP, Institut N\'eel, Grenoble, France}

\def\thefootnote{$*$}\footnotetext{Contact author: olivier.buisson@neel.cnrs.fr}

\date{\today}

\begin{abstract}

Drive-induced unwanted state transitions (DUST) are limiting both for microwave readout and parametric operations of superconducting qubits. Among them, measurement-induced state transitions (MIST) are due to intrinsic resonances described by the readout Hamiltonian. They were previously studied with a qubit linearly coupled to its readout mode, which constitutes the usual readout Hamiltonian. Since MIST can appear even at moderate powers, they limit the readout SNR and the QND readout fidelity. In this work, we study the high-power readout regime in a different transmon readout scheme, implementing a nonlinear coupling called the cos$\varphi$-coupling. This coupling stems from a transmon molecule circuit and has symmetry properties that suppress non-parity-conserving MIST. We succeed in performing multi-state single-shot readout up to the fifth excited state of the transmon, which enables us to  identify leakage pathways from the computational subspace. The measurements indicate that the system is free of MIST up to high powers, with more than 300 photons in the readout mode. 
The MIST can be controllably turned on by breaking the parity symmetry of the coupling using flux-tuning. These experimental results are corroborated by branch analysis and simulations of the classical chaotic dynamics, showing that the cos$\varphi$-coupling is very robust to readout photons compared to the usual transverse coupling.
\end{abstract}

\maketitle

\section{Introduction}
\label{sec:intro}






After decades of development, superconducting qubit architectures \cite{Blais2021,Krantz2019} remain significantly limited by their readout performance \cite{Google2025}. Readout fidelities have only recently surpassed 99.9\% \cite{Spring2024,Kurilovich2025}, which has yet to become commonplace. Furthermore, the standard transmon readout scheme, consisting of the qubit transversely coupled to its readout resonator \cite{Koch2007}, remains overwhelmingly prevalent despite its limitations. In fact, at sufficiently high-photon numbers, required for fast and high-fidelity readout, the measurement loses its quantum non-demolition (QND) character \cite{Sank2016,Khezri2023}.
Thus, high-fidelity readout is currently mostly achieved in the low-power regime \cite{Walter2017,Chen2023,Swiadek2023a,Spring2024}, which significantly limits the signal-to-noise ratio and the readout speed.

Drive-induced unwanted state transitions (DUST) \cite{Dai2025Jun} are one of the main obstacles to fast high-fidelity readout, and, more broadly, to any operation involving a microwave drive in circuit QED. They are associated with an increased probability of transitions between qubit states in the presence of microwave drive photons, including leakage outside of the computational manifold. These transitions can present different physical origins, such as avoided crossings in the readout-drive-dressed spectrum, inelastic scattering and energy exchange with parasitic
two-level systems \cite{Slichter.2012, Thorbeck.2024, Connolly2025Jun,Dai2025Jun}. In particular, transitions which are due to avoided crossings in the dressed spectrum, usually referred to as measurement-induced state transitions (MIST) \cite{Sank2016,Khezri2023}, are intrinsic to the system Hamiltonian and can occur at photon numbers much lower than the empirically formulated critical photon number \cite{Verney2019,Shillito2022,Cohen2023,Dumas2024}. These effects are ubiquitous in transmon-qubit based systems \cite{Sank2016,Khezri2023,Lescanne2019,Fechant2025May,Xia2025Jun}  
and lead to structural instability of the qubit manifold \cite{Cohen2023,Dumas2024}. Recently it was realized for transmon readout that, when the detuning between the qubit and readout resonator is unusually large, such effects are suppressed \cite{Cohen2023,Kurilovich2025,Connolly2025Jun}.

\begin{figure}[t!]
\centering
\includegraphics[width=1\columnwidth]{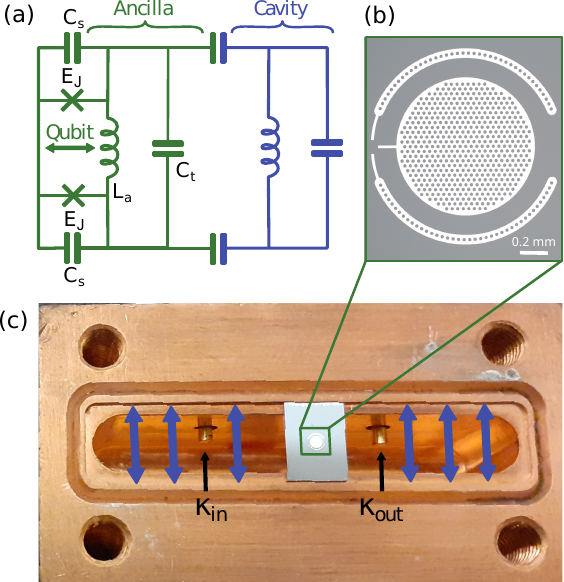}
\caption{\label{fig:circuit} 
(a) Lumped-element circuit for the cos$\varphi$-coupling readout scheme. Transmon molecule in green and cavity in blue. (b) Optical microscope picture of the transmon molecule circuit. (c) Photograph of the transmon molecule sample in its 3D cavity. Blue arrows show the cavity's resonant field direction. The input and output coupling pins are indicated with black arrows.
}
\end{figure}

In this work, we experimentally implement a circuit-QED Hamiltonian that is virtually free 
of measurement-induced transitions by revisiting a transmon readout scheme relying on a purely nonlinear coupling, which we name $\cos\varphi$-coupling. We also provide theoretical arguments for the absence of MIST. The $\cos\varphi$-coupling is implemented by a transmon molecule circuit capacitively coupled to a readout resonator \cite{Diniz2013,Dumur2016,Dassonneville2020,Dassonneville2023} (see \cref{fig:circuit}). The same circuit has been discussed theoretically \cite{Didier2015,Chapple2024} as a means to implement longitudinal readout \cite{Kerman2013Dec,Billangeon2015Mar,Didier2015}. This is part of a larger class of circuits where the qubit-to-resonator coupling is nonlinear \cite{Richer.2017hi8, Ye2024,Hazra2024, Pfeiffer.2024}. Such couplings can lead to high-fidelity readout \cite{Mori2025,Salunkhe2025Jan,Hazra2024,Wang2024Dec}. In particular, the $\cos\varphi$-coupling readout scheme is predicted to be structurally stable up to high photon numbers in the readout resonator \cite{Chapple2024}.

We show here that this robustness is due primarily to the joint effect of a parity symmetry and of detuning, which suppress multi-excitation transitions such as MIST. The matrix elements generating such processes depend on the readout scheme Hamiltonian. In the case of the cos$\varphi$-coupling, its symmetry properties only allow exchange of even numbers of qubit excitations and, most importantly, even numbers of readout photons \cite{Dassonneville2023,Mori2025,Chapple2024}. This parity symmetry forbids many transitions which are otherwise allowed in the case of standard transverse coupling.

The remaining allowed transitions are suppressed by the qubit-to-cavity detuning. In the standard readout scheme, MIST are suppressed exponentially as a function of the ratio $\omega_d/\Wq$, where $\omega_d$ (resp. $\Wq$) is the readout drive (resp. transmon) frequency \cite{Cohen2023,Kurilovich2025,Connolly2025Jun}. Indeed, the strongest allowed MIST employs one readout photon to excite a transition in the transmon and the corresponding matrix element is suppressed for $\omega_d/\Wq\gg 1$. 
For the case of the $\cos \varphi$-coupling readout scheme, due to parity symmetry, 
the lowest relevant spectral collision  occurs at a frequency difference $2\omega_d$. This is associated with an exponential suppression in $2\omega_d/\Wq$ of the matrix elements associated with MIST \cite{Connolly2025Jun}. Thus the cos$\varphi$-coupling is more effective at suppressing the matrix elements leading to MIST than the standard readout scheme.



Here, we experimentally characterize MIST in a $\cos \varphi$-coupled transmon qubit, by measuring the effect of a readout drive pulse of variable power $\bar{n}$ on a prepared qubit state. We are able to precisely determine pairs of states affected by measurement-induced transitions, and the drive amplitudes at which these transitions occur. We observe a remarkable absence of measurement-induced transitions up to 300 photons for initial states $\ket{0}$ and $\ket{1}$. Moreover, we are able to controllably turn on several such measurement-induced transitions by breaking the parity symmetry via a flux bias. We quantitatively confirm these observations by a branch analysis, which consists of studying resonances in the dressed spectrum of the coupled Hamiltonian  \cite{Shillito2022,Dumas2024}. We then compare the benefits of the cos$\varphi$-coupling numerically to the more standard transverse-coupling readout scheme, with branch analysis and classical chaos simulations \cite{Cohen2023,Dumas2024}. 
This allows us to relate the parity-symmetry protection against MIST to the classical structural stability of the $\cos\varphi$-coupling.
 

\section{\texorpdfstring{cos$\varphi$}{cosϕ}-coupling readout}
\label{sec:readout_scheme}

We implement the cos$\varphi$-coupling with a bimodal circuit, called transmon molecule [shown in green in Fig. \ref{fig:circuit}(a)], coupled to a microwave cavity [shown in blue in Fig. \ref{fig:circuit}(a)]. This architecture is introduced in previous works \cite{Dassonneville2020,Dassonneville2023,Mori2025} and results in a transmon (denoted $\textbf{q}$ for  `qubit') cos$\varphi$-coupled to two polariton meter modes. The meter modes stem from the hybridization (details in \cref{Ap:NM}) between the cavity (denoted $\mathbf{c}$) and the transmon molecule ancillary mode (denoted $\mathbf{a}$). This results in a cavity-like polariton (denoted $c$) and an ancilla-like polariton (denoted $a$). In the following, boldface notations are used for bare mode operators.

Each of the two nearly-linear polaritons can be used for qubit readout since they are $\cos\varphi$-coupled to the qubit and are also coupled to the exterior through the cavity. In practice, we choose the more linear cavity-like polariton as the readout mode. As the two polaritons are far detuned, we can assume that the ancilla-like polariton remains in its ground state throughout the readout protocol. The effective Hamiltonian describing only the two remaining active degrees of freedom (derived in \cref{apx:theory}) is given by:

\begin{align}
\label{eq:2mode_ham}
\begin{split}
    \hat{H} =& 4 \ECq (\bnq-n_g)^2-2 E_J \cos \left(\bphq\right) + \hbar\wc \hc^{\dagger} \hc \\
    &- 2 \bar{E}_J\left[\cos \left(\bphq\right)-1\right] \\
         &\hspace{30pt}\left\{\cos \left[\varphi_c (\hat{c}+\hat{c}^\dagger) + \bar{\varphi}_\text{ext} + \tilde{\eta}(t) \right]-1\right\},
\end{split}
\end{align}
where the qubit mode is defined by its charging (resp. Josephson) energy $\ECq$ (resp. $2E_J$) and its charge number (resp. superconducting phase) operator $\bnq$ (resp. $\bphq$). The readout polariton is defined by its frequency $\wc$, annihilation operator $\hc$ and zero-point phase fluctuations $\varphi_c$. Small corrections are made to the Josephson energy $\bar{E}_J \approx E_J$ due to the zero-point fluctuations of the ancilla mode. 
\Cref{eq:2mode_ham} realizes an effective $\cos\varphi$-coupling between the transmon qubit mode and the readout cavity mode. The flux dependence appears in $\bar{\varphi}_\text{ext}$. In practice, the flux bias is set to an integer multiple flux quanta, resulting in $\bar{\varphi}_\text{ext}=0$. The readout drive dependence is encoded in $\tilde{\eta}(t)$, which is vanishing at zero drive amplitude. Hereafter, when in the context of the two-mode Hamiltonian \cref{eq:2mode_ham}, we shall refer succinctly to the `cavity polariton' as `the cavity'.


At vanishing flux and drive, the cos$\varphi$-coupling couples the qubit $\cos(\bphq)$ operator to the cavity $\cos[\varphi_c(\hc+\hc^\dagger)]$ operator. When approximating the transmon qubit as a weakly anharmonic oscillator $\bphq = \bm{\varphi}_q (\hbq+\hbq^\dagger)$, with $\bm{\varphi}_q$ its zero-point phase fluctuations, we can Taylor expand the cosine potentials around their minima assuming that zero-point fluctuations of phase are small $\bm{\varphi}_q,\varphi_c \ll 1$. 
The dominant term of this expanded form is the cross-Kerr interaction $\chi_{qc} \hbq^\dagger \hbq \hat{c}^\dagger \hat{c}$, with dispersive shift  $\chi_{qc} = -2 \bar{E}_J \bm{\varphi}_q^2 \varphi_c^2/4$. Note that this cross-Kerr coupling mediates QND readout \cite{Dassonneville2020,Dassonneville2023,Mori2025} and does not arise from a perturbative  approximation to a linear capacitive coupling as in the case of standard transverse readout. It is also independent of the qubit-to-cavity detuning. 

In its exact form, the cos$\varphi$-coupling obeys parity symmetry due to the matrix elements of the underlying operators $\cos(\bphq)$ and $\cos[\varphi_c(\hc+\hc^\dagger)]$. We show in the following sections that this results in a readout scheme which is more robust to readout photons than the transverse coupling. Indeed the symmetry properties forbid all non-parity-conserving transitions, resulting in a much less allowed transitions in the readout-drive-dressed qubit system. Furthermore, the suppression of MIST for large qubit-to-cavity detunings, demonstrated for transversely coupled transmons \cite{Kurilovich2025,Connolly2025Jun}, is more effective in the case of a $\cos\varphi$-coupled transmon. These symmetry properties also result in a structural stability of the cos$\varphi$-coupling, as pointed out recently in a theoretical study \cite{Chapple2024}. 




To demonstrate this new readout scheme, we have fabricated  the transmon molecule circuit with aluminum on a silicon chip [see \cref{fig:circuit}(b,c)] and placed it in a 3D microwave resonator (see \cref{apx:samplefab} and \cref{apx:setup} for details on the fabrication and microwave setup). 
We  have determined the sample parameters from fits of spectroscopic data (see \cref{apx:fit}), as shown in \cref{tab:main_params_only}. In a previous work, this sample also displayed a readout fidelity of $99.21\%$, measured with a power corresponding to $\bar{n}=90$ readout photons and an integration time of $\SI{400}{\nano\second}$ \cite{Mori2025}. This underlines the robustness of the cos$\varphi$-coupling readout scheme to large numbers of photons.

\begin{table}
\centering
\begin{tabular}{c@{\hspace{2em}}c} 
\toprule
Parameter name & Value \\
\toprule
$\ECq/h$ & \SI{0.0734}{\giga\hertz} \\
$2E_J/h$ & \SI{7.92}{\giga\hertz} \\
$\Wq/2\pi$ & \SI{2.0687}{\giga\hertz} \\
$\omega_c/2\pi$ & \SI{7.294}{\giga\hertz} \\
$\balq/2\pi$ & \SI{-81.4}{\mega\hertz} \\
$\alpha_c/2\pi$ & \SI{-0.0155}{\mega\hertz} \\
$\kappa_c/2\pi$ & \SI{17.2}{\mega\hertz} \\
$\chi_{qc}/2\pi$ & \SI{-2.02}{\mega\hertz} \\
\bottomrule
\end{tabular}
\caption{Main sample parameters. Note that the effective Josephson energy of the transmon mode is $2E_J$ instead of $E_J$. }
\label{tab:main_params_only}
\end{table}


\begin{figure*}[t]
\centering
\includegraphics[width=1.9\columnwidth]{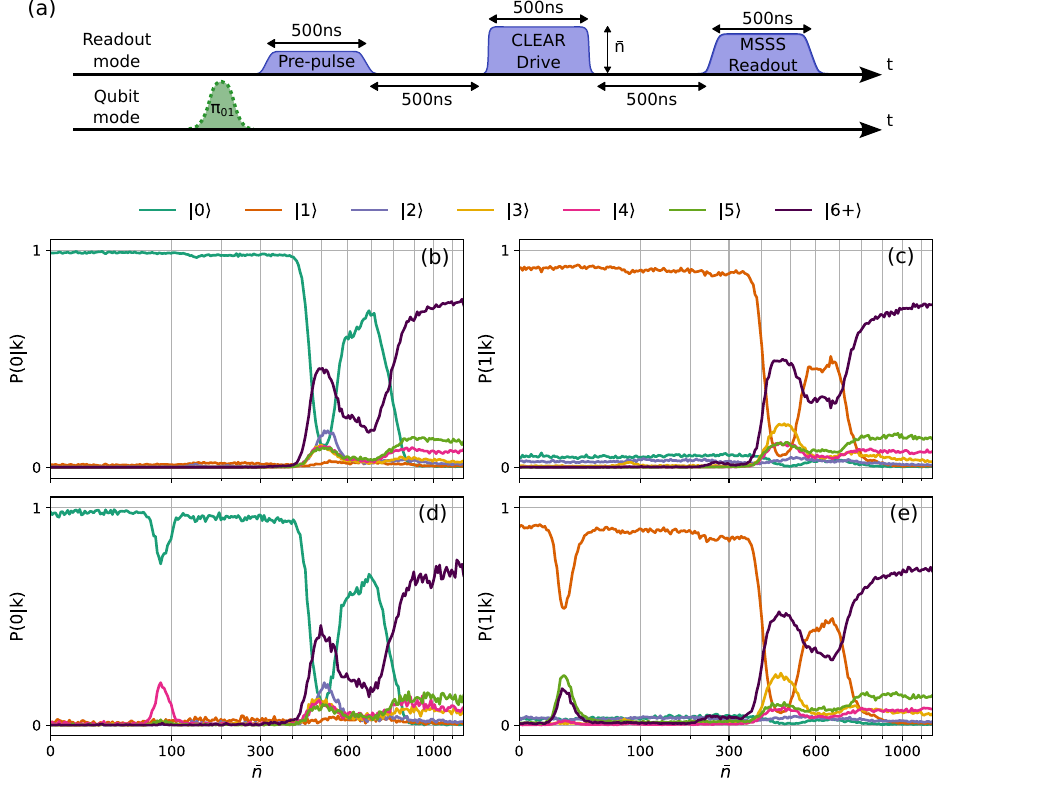}
\caption{\label{fig:mist_meas} \textit{MIST measurements at zero and non-zero flux}. 
(a) Pulse sequence with optional $\pi_{01}$-pulse to prepare $\ket{1}$. (b)  Measurement at zero flux with $\left|0\right\rangle$ initially prepared. Colored lines correspond to probabilities of ending in a given transmon state $\left|k\right\rangle$ as function of the drive power $\bar{n}$. Further panels: analogous measurements for (c)  zero flux with $\left|1\right\rangle$ initially prepared, (d) non-zero flux bias $\Phi_\text{ext}=-0.04\Phi_0$ with $\left|0\right\rangle$ initially prepared, and (e) non-zero flux bias $\Phi_\text{ext}=-0.04\Phi_0$ with $\left|1\right\rangle$ initially prepared.  }
\end{figure*}

\section{MIST measurements and simulations}
\label{sec:mist_meas}

\begin{figure*}[!ht]
\centering
\includegraphics[width=1.9\columnwidth]{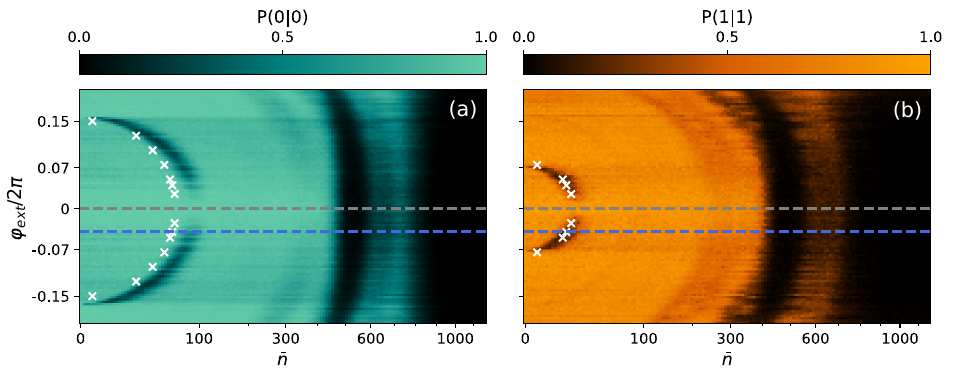}
\caption{\label{fig:mist_vs_flux} \textit{MIST measurements as function of flux}. 
(a) Colorplot of the probability of the transmon remaining in state $\left|0\right\rangle$ in the presence of a cavity drive, as a function of the drive power $\bar{n}$ and the reduced flux bias applied to the sample. Horizontal grey and blue dashed lines indicate the flux biases $\Phi_\text{ext}=0\Phi_0$ and $\Phi_\text{ext}=-0.04\Phi_0$, respectively. White crosses correspond to the position of the $\ket{0}-\ket{4}$ MIST predicted by branch analysis. (b) Colorplot of the probability of the transmon remaining in state $\left|1\right\rangle$ in the presence of a cavity drive, as a function of the drive power $\bar{n}$ and the reduced flux bias applied to the sample. Horizontal grey and blue dashed lines indicate the flux biases $\Phi_\text{ext}=0\Phi_0$ and $\Phi_\text{ext}=-0.04\Phi_0$, respectively. White crosses correspond to the position of the $\ket{1}-\ket{5}$ MIST predicted by branch analysis. }
\end{figure*}



We directly probe MIST with a protocol similar to the one in Ref. \cite{Sank2016,Khezri2023}. The pulse sequence in Fig. \ref{fig:mist_meas}(a) involves preparing the qubit state with an optional $\pi$-pulse and a lower power pre-pulse for post-selection, then populating the readout mode with a drive of variable power $\bar{n}$ and finally measuring the final state of the qubit. In the following, the term ``qubit state'' is used generally for qubit mode states $\ket{k}$ even if $k>1$. Note that the variable-power drive has a CLEAR pulse shape \cite{McClure2016}, in order to accelerate the ring-up and ring-down (details in \cref{apx:clear}). The durations of the pre-pulse, of the variable-power drive, and of the final readout pulse  are each \SI{500}{\nano\second}. This protocol provides a straightforward study of the transitions induced on the qubit as a function of the number of readout photons $\bar{n}$, which we calibrate using the qubit AC Stark shift (see \cref{apx:nphcalib}). 

As MIST are expected to cause leakage out of the computational space, we developed a multi-state single shot (MSSS) readout by tuning sample parameters to differentiate as many states as possible in the IQ plane.
To achieve this readout, we fix a ratio between the cross-Kerr associated with the qubit transition and the cavity decay rate of $\chi_{qc}/\kappa_c= 0.12$, leading to a very regular dispersive shift for qubit states $\ket{0}$ through $\ket{5}$, all of which are then identifiable from single-shot readout. Furthermore, all higher states, denoted hereafter $\ket{6+}$, are guaranteed to not overlap with the lower identifiable states. We postselect out  measured points which are ambiguous and cannot be easily attributed to a given state. Details on the MSSS readout are given in \cref{apx:iq_plane}. 



Remarkably, we find that, at the symmetric zero-flux point, MIST involving the computational states  $\ket{0}$ and $\ket{1}$ are absent up to $\bar{n} \approx 300$ photons.  
To analyze the QND character of the measurement, we plot the conditional probabilities $P(i|f)$, for initial post-selected state $i$ and measured final state $f$, as a function of $\bar{n}$. Fig. \ref{fig:mist_meas} (b,c) shows the resulting plot for prepared state $\ket{i}=\ket{0}$ or $\ket{1}$, respectively, with different colored lines for each possible final state $f \in \{0,1,2,3,4,5,6+\}$. 
For both prepared states, the features of the MIST measurement are quite similar. First, the probability of conserving the initial state $P(i|i)$ stays relatively constant in a plateau-like regime. The plateau when preparing $\ket{0}$ is stable around 97\% up to 373 photons. For a prepared $\ket{1}$-state, the plateau stays around 90\% up to 326 photons. These plateaus indicate that the qubit state is quite resilient to readout photons up to around $\bar{n}=300$. Note that the plateau for $P(1|1)$ is lower than for $P(0|0)$ due to additional preparation errors. 

After about 300 photons, these plateaus are interrupted by strong non-QND features, subsequently limiting the probabilities $P(i|i)$ to less than 70\%. The advent of these very high power non-QND phenomena is around the expected power where the non-linearity of the readout mode is expected to become apparent. This could be evidence of the emergence of the bistability regime, estimated around $\bar{n}_\text{bifurc} = 231$ photons \cite{Mori2025,Dassonneville2023,Ong2011}, or it could signal that a Taylor expansion of the cos$\varphi$-coupling is invalid around $\bar{n}_{\text{low}\varphi}= 286$ photons \cite{Mori2025}, or, furthermore, it could correspond to inelastic scattering in the system
\cite{Connolly2025Jun,Dai2025Jun}.



However, the regular plateau-like regime up to 300 photons is a signature of the $\cos\varphi$-coupling symmetry properties. To highlight this feature, we now tune the flux bias away from the sweetspot at zero flux. Focusing on  $\Phi_\text{ext} = -0.04\Phi_0$, we find that certain MIST are activated, see Fig. \ref{fig:mist_meas}(d,e). New  features appear, interrupting the plateau regime long before 300 photons. When preparing $\ket{0}$, the probability $P(0|0)$ presents a new resonant dip at 81 photons, accompanied by a symmetric peak in $P(0|4)$. For a prepared $\ket{1}$-state, a new dip in $P(1|1)$ appears at 13 photons, accompanied mainly by a peak in $P(1|5)$ and a slightly lower peak in $P(1|6+)$. We associate these features to measurement-induced hybridization of qubit states $\ket{0}$ and $\ket{4}$, and $\ket{1}$ and $\ket{5}$, respectively, which results in coherent population transfer as the cavity population rings up. These two features will be referred to as the $\ket{0}-\ket{4}$ and the $\ket{1}-\ket{5}$ MIST hereafter.

We now perform a more complete flux sweep in the range $\pm 0.2\Phi_0$, while initializing the qubit in either $\ket{0}$ and $\ket{1}$.  In Fig. \ref{fig:mist_vs_flux} we plot 
$P(i|i)$ as function of flux and number of photons $\bar{n}$. We confirm that the plateau-like regime is specific to the zero-flux point. As soon as the flux bias is off this sweetspot, features showing as darker lines [suppression of $P(i|i)$] appear due to  the $\ket{0}-\ket{4}$ and the $\ket{1}-\ket{5}$ MIST. The positions of these resonant features shift towards lower powers as the flux increases away from the sweetspot until, eventually, they reach 0 photons and fully disappear. The $\ket{0}-\ket{4}$ MIST disappears around $\pm 0.16\Phi$ and the $\ket{1}-\ket{5}$ disappears at $\pm 0.07\Phi_0$. These points coincide with resonances in the dressed spectrum between the respective qubit transition and the cavity frequency (\cref{apx:theory}).  

We can quantitatively model the situation above by means of a branch analysis \cite{Shillito2022,Dumas2024} of the undriven two-mode Hamiltonian, \cref{eq:2mode_ham} with $\tilde{\eta}(t)=0$ (for details see \cref{Ap:BA}). In \cref{fig:mist_vs_flux} , crosses mark the position of predicted avoided crossings in the dressed spectrum corresponding to the $\ket{0}-\ket{4}$ and $\ket{1}-\ket{5}$ resonances, respectively. The branch analysis predicts, in line with the experimental data, a complete absence of MIST at zero flux. 
This shows good quantitative agreement between the numerics and experimental data.

\begin{figure*}[t!]
    \centering
    \includegraphics[width=1.8\columnwidth]{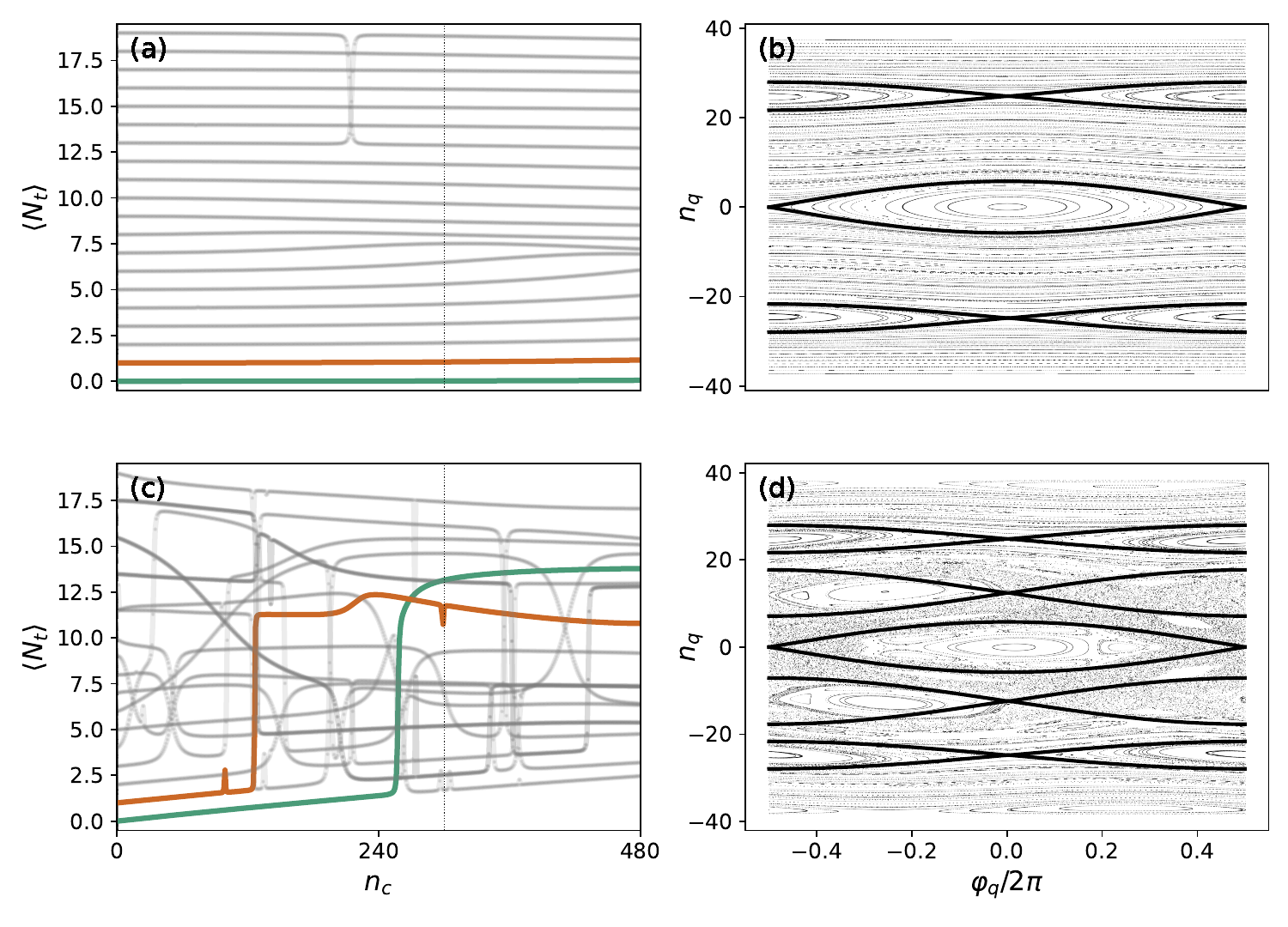}
    \caption{\textit{Comparison between cos$\varphi$-coupling and transverse coupling}. 
    (a) and (c) Branch analysis for a cos$\varphi$-coupled transmon and an equivalent transversely coupled transmon [$j=0$ and $j=1$ branches highlighted with color conventions of \cref{fig:mist_meas}(b)-(e)]. (b) and (d): Corresponding Poincaré sections at $\bar{n}=300$ [vertical dashed line in (a),(c)] with separatrices corresponding to  nonlinear resonances. Odd harmonics of the drive are absent for the cos$\varphi$-coupling (b), rendering it robust to the formation of chaotic regions since separatrices corresponding to the zeroth and second harmonics are well separated. In the case of a transversely coupled transmon (d),  resonances corresponding to the first harmonic of the drive almost overlap along the $n_q$-axis with the zeroth harmonic, which is consistent with extended chaotic regions that are no longer confined to the neighborhood of the separatrices  \cite{Lichtenberg,Chirikov1979May} (see \cref{Ap:Stability}).}
    \label{fig:BrPo}
\end{figure*}


The model used for the branch analysis allows us to explain the relevant measurement-induced transitions by invoking the symmetry properties of the $\cos\varphi$-coupling. As previously mentioned, these properties stem from the matrix elements of the operators which are coupled together. When the flux bias is not an integer number of flux quanta, additional matrix elements emerge in the coupling Hamiltonian. This can be made explicit in the flux-dependent two-mode Hamiltonian \cref{eq:2mode_ham}, written as follows:
\begin{align}
\label{eq:2mode_ham_nonzeroflux}
\begin{split}
    \hat{H}_{qc} &= 4 \ECq \bnq^2-2 E_J \cos \left(\bphq\right) + \hbar\wc \hc^{\dagger} \hc \\
    &-2\bar{E}_J \left[\cos(\bphq)-1\right] \left[\cos\varphi_c(\hc+\hc^\dagger)\cos\bar{\varphi}_\text{ext}-1\right] \\
    & +2\bar{E}_J \left[\cos(\bphq)-1\right] \left[\sin\varphi_c(\hc+\hc^\dagger)\sin\bar{\varphi}_\text{ext}\right].
\end{split}
\end{align}
At non-zero flux bias, the last line of Eq. (\ref{eq:2mode_ham_nonzeroflux}) will lead to additional matrix elements and breaking the cavity parity-symmetry. To see this explicitly,  Taylor-expanding this line results in terms $\hbq^{2n}(\hc^\dag)^{2m-1} + \rm{H.c.}$, with nonnegative integers $n,m$. They notably give rise to terms $\hbq^4\hc^\dag$ and $(\hbq^\dag)^4\hc$, which mediate the $\ket{0}-\ket{4}$ and $\ket{1}-\ket{5}$ MIST. At zero flux $\bar{\varphi}_\text{ext}=0$, these contributions are absent from the Hamiltonian, and the transitions are forbidden.

\section{Comparison to the transverse coupling}
\label{sec:comparison}

Having elucidated the nature of MIST in the transmon molecule circuit, we now show that the $\cos \varphi$-coupling (at zero flux bias) is more robust against MIST than a transversely coupled transmon, which is typically susceptible to ionization \cite{Shillito2022,Cohen2023,Dumas2024}. To this end we derive, for the two-mode cos$\varphi$-coupling model \cref{eq:2mode_ham}, a transversely coupled circuit-QED model \cite{Shillito2022}
\begin{align}
    \begin{split}
        \label{eq:2mode_ham_transverse}
    \hat{H}' =& 4 \ECq (\bnq-n_g)^2-2 E_J \cos \left(\bphq\right) \\
    &+ \hbar\wc \hc^\dagger \hc - i \hbar\varepsilon_d(t) (\hc-\hc^\dagger) - i \hbar g_{qc} \bnq (\hc-\hc^\dagger).
    \end{split}
\end{align} 
We enforce that the dispersive interaction and qubit anharmonicites are the same in the two models, amounting to choosing a transverse coupling strength $g_{qc} = |\varphi_c| \omega_c$. As such, the standard transversely coupled transmon in \cref{eq:2mode_ham_transverse} has the same linear response as the cos$\varphi$-coupling in \cref{eq:2mode_ham} under a readout drive (for details, see \cref{apx:eq_transverse}). The two models display the same AC Stark shift in the presence of readout photons and also have equal cross-Kerr interaction. This AC Stark shift is also consistent with the experimental photon number calibration (see \cref{apxfig:nph_calib}).

We compare the two types of coupling with the same kind of branch analysis as above (for details, see \cref{Ap:Stability}). This consists of diagonalizing the coupled Hamiltonian and employing an adiabatic state-labeling procedure to identify the dressed eigenstates $\ket{\overline{j,n_c}}$, with $j$ (resp. $n_c$) the number of transmon (resp. cavity) excitations. The bare basis transmon occupation number
\begin{align}
    \label{eq:nt}
    \hat{N}_t = \sum_{j \geq 0} j \ket{j}\bra{j} \otimes \hat{\mathbb{1}}_c,
\end{align} 
where $\ket{j}$ are the eigenstates of the bare transmon mode and $\hat{\mathbb{1}}_c$ is the identity operator over the cavity Hilbert space, is a convenient indicator of avoided crossings in the dressed spectrum of the undriven Hamiltonian, induced by the qubit-to-cavity coupling \cite{Shillito2022,Cohen2023,Dumas2024}. 


Focusing first on the cos$\varphi$-coupling \cref{eq:2mode_ham}, in \cref{fig:BrPo}(a) we show the expectation value of the transmon excitation number in \cref{eq:nt}, \ie $\langle N_t \rangle \equiv \langle \overline{j, n_c} | \hat{N}_t |  \overline{j, n_c} \rangle$ versus $n_c$ on the $x$-axis, and colors encoding the excitation number of the transmon, $j$, obeying the same conventions as \cref{fig:mist_meas}. The absence of any crossings of the curves corresponding to the states in the computational manifold (green for $j=0$ and orange for $j=1$) indicates that the corresponding branches of the cavity mode, $\ket{\overline{0, n_c}}$ and $\ket{\overline{1, n_c}}$, have no discernible avoided crossings with other branches, which precludes MIST involving the two states of the computational manifold. 

Turning to the equivalent two-mode transversely coupled Hamiltonian $\hH'$ in \cref{eq:2mode_ham_transverse}, we note that the branches corresponding to the computational manifold cross with higher-lying states [\cref{fig:BrPo}(c)]. The two computational manifold states can come into resonance with chaotic states \cite{Cohen2023,Dumas2024}. We note that the curves in \cref{fig:BrPo}(c) have marked slopes, which indicates significant dressing of the eigenvectors due to the drive, as opposed to the branches simulated for cos$\varphi$-coupling. In spectroscopy, however, we have checked that the transmon 0-1 transition frequency and its linear AC Stark shift response are the same for both types of coupling. This simulated AC Stark shift is also consistent with experiment (\cref{apx:nphcalib}). Thus it appears that the $\cos\varphi$-coupled transmon mode can be AC Stark shifted by an amount substantially exceeding its anharmonicity without being affected by MIST.


\section{Structural stability}
\label{sec:chaos}

Beyond the study of resonant effects such as MIST, the cos$\varphi$-coupling and transverse coupling can be compared with respect to their overall stability towards the onset of chaotic motion in the presence of a strong classical drive. We first derive a semi-classical model for each readout scheme and further use it for classical simulations to obtain Poincaré sections \cite{Cohen2023,Dumas2024}. When driving the transmon through a transverse coupling, the semi-classical Hamiltonian describing the transmon mode takes the form 
\begin{align}
\label{eq:semiclass_transverse}
\begin{split}
    \hat{H}_{sc}' &= 4 \ECq \bnq^2 - 2E_J J_0\left(\tilde{\eta}_{t,0} \right) \cos\left(\bphq\right) \\
    &- 2E_J \sin\left(\bphq\right)\sum_{n\in\mathbb{N}} J_{2n-1}\left(\tilde{\eta}_{t,0} \right) \sin\left[\left(2n-1\right)\wdr t\right] \\
    &- 2E_J \cos\left(\bphq\right)\sum_{n\in\mathbb{N}^*} J_{2n}\left(\tilde{\eta}_{t,0} \right) \cos\left(2n\wdr t\right),
\end{split}
\end{align}
where $J_n$ are the Bessel functions of the first kind \cite{Abramowitz1964}, $\tilde{\eta}_{t,0} = 2g_{qc}\sqrt{\bar{n}}/\omega_d$ and $\bar{n}$ is the power of the drive, in average number of photons. Note that the drive terms are split between odd harmonics, coupled to the $\sin(\bphq)$ operator, and even harmonics, coupled to the $\cos(\bphq)$ operator. Furthermore, the $0^\textit{th}$ harmonic is separated from the sum since it does not induce transitions. In the presence of a drive, $J_0(\tilde{\eta}_{t,0})$ encodes the AC Stark shift on the transmon spectrum.

In the case of a transmon driven through a cos$\varphi$-coupling (at zero flux bias), the semi-classical Hamiltonian takes the following form (derived in Appendix \ref{Ap:Stability}):
\begin{align}
\label{eq:semiclass_cosphi}
\begin{split}
    \hat{H}_{sc} &= 4 \ECq \bnq^2 - 2E_J J_0\left(\tilde{\eta}_0 \right) \cos\left(\bphq\right) \\
    &-2E_J \cos\left(\bphq\right)\sum_{n\in\mathbb{N}^*} (-1)^n J_{2n}\left(\tilde{\eta}_0 \right) \cos\left(2n\wdr t\right),
\end{split}
\end{align}
with the dimensionless drive magnitude $\tilde{\eta}_0 = 2\varphi_c\sqrt{\bar{n}}$. In contrast to the transverse coupling, the parity-symmetry properties of the cos$\varphi$-coupling result in only even-numbered drive harmonics. Note that to ensure equal AC Stark shifts we have $\tilde{\eta}_0 = \tilde{\eta}_{t,0}$, as discussed in \cref{sec:comparison}.

To study the structural stability of each type of coupling, \cref{eq:semiclass_transverse} and \cref{eq:semiclass_cosphi}, we take the classical limit and generate the corresponding Poincaré sections \cite{Zaslavskii1991Apr}, corresponding to plotting a point at phase-space coordinates $\left(\bm{\varphi}_q(n T),\bm{n}_q(n T)\right)$ at each integer multiple of the driving period $nT$, for a chosen set of initial conditions. We show these in \cref{fig:BrPo}(b,d) for a drive power corresponding to $\bar{n} = 300$ photons. 
In regular regions of the phase space, the sequence of points obtained after every period coalesces into closed curves, either consisting of unbounded trajectories (curves close over the periodic boundary condition at $\bm{\varphi}_q = \pm \pi$) or bounded oscillations. The central regular island located around $\bm{n}_q =0$ and $\bm{\varphi}_q=0$, containing bounded trajectories with closed ellipsoidal shape, hosts the states belonging to the transmon computational manifold \cite{Cohen2023}. This regular region corresponds to the so-called principal resonance \cite{Chirikov1979May,Lichtenberg}, generated by the zeroth harmonic of the drive frequency. At zero drive amplitude, it corresponds to the bounded orbits of the  undriven oscillator. The applied drive generates, however, additional resonances for each harmonic of the drive frequency (for a more detailed discussion, see \cref{Ap:Stability}). These are regions of phase space where regular bounded motion  can occur, which are centered at $\bm{n}_q \neq 0$ and whose width along the Cooper-pair number $\bm{n}_q$-axis depends on drive power. These regions can be delimited approximately by so-called separatrices, represented as black lines  (see \cref{Ap:Stability}).

Furthermore, the presence of a drive can give rise to chaotic dynamics  \cite{Cohen2023,Dumas2024}. For initial conditions within the so-called chaotic layer, trajectories on the Poincaré sections no longer coalesce into closed one-dimensional curves, but instead densely fill regions of the two-dimensional phase space $(\bm{\varphi}_q,\bm{n}_q)$. Such regions are clearly visible for transverse coupling \cref{fig:BrPo}(d), whereas in the case of the $\cos\varphi$ coupling, they are practically invisible. 

To understand the absence of chaotic regions in the Poincaré sections, we invoke a heuristic qualitative criterion due to Chirikov \cite{Chirikov1979May,Lichtenberg} (see \cref{Ap:Stability}). According to this, chaotic motion can occur whenever two nonlinear resonances come close to overlapping along the Cooper-pair axis $\bm{n}_q$ (trajectories initialized near these two resonances are not necessarily deterministic). Importantly, we observe twice less nonlinear-resonance separatrices  for the cos$\varphi$-coupling compared to the transverse coupling, due to the absence of odd-ordered harmonics in the former [compare the semiclassical Hamiltonian \cref{eq:semiclass_cosphi} to \cref{eq:semiclass_transverse}]. This relative sparsity of the nonlinear resonances   
renders the $\cos\varphi$ coupling significantly more robust toward the possibility of chaotic dynamics. Indeed, in this case, the chaotic  regions are expected to remain exponentially localized in the neighborhood of the separatrices \cite{Zaslavskii1991Apr,Bubner1991Feb,Cohen2023}, as the transmon mode is effectively driven by the \textit{second} harmonic of the readout drive, a far off-resonant tone, $2\omega_c/\bm{\omega}_q \approx 7$, a fact which was exploited in standard transversely-coupled transmons recently  \cite{Kurilovich2025}. 
Note, however, that in implementations where the second harmonic of the drive is closer to the qubit plasma frequency, a chaotic layer can visibly  develop for the cos$\varphi$-coupling as well \cite{Chapple2024} (not shown here).

\section{Conclusion}
\label{sec:conclusion}

In summary, we have implemented transmon readout with a purely non-linear coupling, the cos$\varphi$-coupling, which is free of measurement-induced transitions up to high powers. Using optimized multi-state single-shot readout, we have been able to comprehensively study the onset of MIST in the system. At zero flux, consistent with the symmetry properties of the $\cos\varphi$-coupling, the qubit computational manifold is undisturbed by readout photons up to $\bar{n}\approx 300$ photons, beyond which we believe the QND aspect of the readout is interrupted by the cavity bistability threshold, estimated around 231 photons, or by inelastic scattering \cite{Connolly2025Jun,Dai2025Jun}. When the flux symmetry is broken, the measurements reveal the appearance of additional MIST, which can be traced back to matrix elements that are otherwise completely canceled at vanishing flux bias due to the cos$\varphi$-coupling parity symmetry. We have been able to quantitatively predict the underlying spectral resonances responsible for these transitions using numerical simulations. 

Our proof-of-principle implementation shows that the $\cos\varphi$-coupling readout scheme is more robust to MIST than transmon readout with transverse coupling. The central ingredient for this stability is its parity symmetry, which forbids any transitions with an odd number of transmon or cavity excitations, otherwise allowed in the standard readout scheme. Furthermore, the remaining transitions are exponentially suppressed in the large qubit-to-cavity detuning regime. This is due to an exponential suppression of MIST in far-detuned readout, 
which is also observed in transversely coupled transmons \cite{Kurilovich2025,Connolly2025Jun,Dai2025Jun}, but is more effective for cos$\varphi$-coupling. Indeed, for a drive of frequency $\omega_d$, the parity-symmetry only allows even-numbered drive harmonics, meaning that the lowest harmonic which can affect the transmon is at $2\omega_d$ instead of $\omega_d$. This also results in an increased immunity to chaotic effects in the presence of very high readout powers, at which the transmon mode would ionize in transverse readout. Beyond simply suppressing MIST, the cos$\varphi$-coupling shows better structural stability than the standard transmon readout scheme, in the presence of strong drives. 
This robustness to off-resonant parametric drives establishes the cos$\varphi$-coupling as a compelling alternative for high-fidelity qubit readout.




\begin{acknowledgments}

The authors thank J. Cohen, R. Dassonneville, B. Huard, M. Mirrahimi, and P. Leek for insightful discussions.

This work benefited from a French government grant managed by the ANR agency under the ‘France 2030 plan’, with reference ANR-22-PETQ-0003 and BPI France AD ASTRA. O.B., C.M., F.D’E. and AP acknowledge support from ANR OCTAVES (ANR-21-CE47-0007). 

\end{acknowledgments}

\appendix

\section{Theoretical model for the \texorpdfstring{cos$\varphi$}{cosϕ}-coupling readout scheme}
\label{apx:theory}
We obtain by quantizing the circuit in \cref{fig:circuit} the following model Hamiltonian \cite{Dassonneville2020,Dassonneville2023,Mori2025}
\begin{align}
    \hH = \hH_\mathbf{q} + \hH_\mathbf{a} + \hH_\mathbf{c} + \hH_\mathbf{qa} + \hH_\mathbf{ac}, \label{eq:Htot}
\end{align}
where 
\begin{align}
    \begin{split}\label{eq:selfH}
    \hH_\mathbf{q} &=  4 \ECq(\bnq - n_g)^2-2 E_J \cos \left(\bphq\right),  \\
    \hH_\mathbf{a} &= 4 \ECa \bna^2 -2 E_J\left[\cos \left(\bpha\right)-\frac{L_J}{L_a} \left(\bpha - \half\phx\right)^2\right], \\ 
    \hH_\mathbf{c} &= \hbar \Wc \hbc^{\dagger} \hbc + \hbar\eps_d (t) (-i)(\hbc-\hbc^\dagger)
    \end{split}
\end{align}
are the self-Hamiltonians of the transmon, ancilla, and cavity modes, and their couplings are given by
\begin{align}
    \begin{split}
        \hH_\mathbf{qa} &= -2 E_J\left[\cos \left(\bphq\right)-1\right]\left[\cos \left(\bpha\right)-1\right],\\
        \hH_\mathbf{ac} &= - \hbar g_{a c}\left(\hba - \hba^{\dagger}\right)\left(\hbc- \hbc^{\dagger}\right).
    \end{split}
\end{align}
We note that the inductance $L_a$ originates from a SQUID loop and has a weak flux dependence $L_a = L_a(\varphi_\text{ext}) / L_a(0)/|\cos(\pi \varphi_\text{ext} / 28)|$, where 28 is the ratio between the area of the large transmon molecule superconducting loop and the area of a single SQUID loop. The readout drive power $\epsilon_d$ and frequency $\omega_d$ are encoded in $\varepsilon_d(t) = \epsilon_d \sin(\omega_dt)$. 

Since $L_J/L_a\gg1$ for our sample parameters, the ancilla mode anharmonicity is low. By linearizing it, we perform a change of basis on the Hamiltonian \cref{eq:Htot} that describes as closely as possible the readout mode used in the experiment. This nearly linear mode is the cavity-like polariton that is obtained upon performing a Bogoliubov transformation to eliminate the linear capacitive coupling $g_{ac}$. The final result of this section is the Hamiltonian given in \cref{eq:HsumAgain,eq:HintAgain}.  

The Hamiltonian is completely rewritten in the polariton basis
\begin{align}
    \begin{split} \label{eq:Hsum}
    \hH &= \hH_\mathbf{q} + \hH_a + \hH_c + \hH_{\mathbf{q}ac} + \hH_{ac},
    \end{split}
\end{align}
with $\hH_\mathbf{q}$ as defined before, and now (see \cref{Ap:NM} for the definitions of the coefficients $v_{ij}$ and $u_{ij}$, $i,j = a,c$ used below)
\begin{align}
    \begin{split}\label{eq:HaHc}
        \hH_a =& \hbar \wa \ha^\dagger \ha -i  v_{ca} \hbar\eps_d (t) (\ha-\ha^\dagger) -\hbar\wa\alpha_a (\ha+\ha^\dagger), \\  
        \hH_c =& \hbar \wc \hc^\dagger \hc -i v_{cc} \hbar\eps_d (t) (\hc-\hc^\dagger) -\hbar\wc\alpha_c (\hc + \hc^\dagger)    
    \end{split}
\end{align}
with $\alpha_i =  \varphi_\text{ext} \bm{\varphi}_a  u_{ai} (L_J/L_a)(2E_J/\hbar\omega_i)$ for $i=a,c$, where $\varphi_\text{ext}$ is the reduced applied flux and $\bm{\varphi}_a = [\ECa/(E_J(1+2L_J/L_a))]^{1/4}$ encodes phase zero-point fluctuations for the ancilla mode. The remaining coupling between ancilla and cavity is now restricted to the Josephson terms
\begin{align}
    \begin{split} \label{eq:Hint}
        \hH_{ac} &= - 2 E_J \cos \left(\bpha\right) \\
        & =- 2 E_J \cos \Big\{\bm{\varphi}_a\left[ u_{aa} (\ha+\ha^\dagger) + u_{ac} (\hc + \hc^\dagger)  \right] \Big\},\\
        \hH_{\mathbf{q}ac} &=  -2 E_J\left[\cos \left(\bphq\right)-1\right]\left[\cos \left(\bpha\right)-1\right] \\
        &= -2 E_J\left[\cos \left(\bphq\right)-1\right] \times \\
        &\Big(\cos\Big\{\bm{\varphi}_a\left[ u_{aa} (\ha+\ha^\dagger) + u_{ac} (\hc + \hc^\dagger)  \right]\Big\}-1\Big).
    \end{split}
\end{align}
This is now a complete rewriting of the original Hamiltonian. 

In the next step, we perform a time-independent displacement transformation to remove the linear flux term in \cref{eq:HaHc}. Taking $\hat{D}_a = e^{\alpha_a \ha^\dagger - \alpha_a^* \ha}$ and $\hat{D}_c = e^{\alpha_c \hc^\dagger - \alpha_c^* \hc}$, we have, up to scalars, $D^\dagger_a\left[  \ha^\dagger \ha + \alpha_a (\ha+\ha^\dagger) \right] D_a =  \ha^\dagger \ha$ and the analogous equation for mode $c$. This procedure moves the flux term from \cref{eq:HaHc} into \cref{eq:Hint},
\begin{align}
    \begin{split}
        \hH_{ac} &=- 2 E_J \cos\Big\{\bm{\varphi}_a\big[ u_{aa} (\ha+\ha^\dagger + 2 \alpha_a )\\
        &+ u_{ac} (\hc + \hc^\dagger + 2 \alpha_c )  \big]\Big\},\\
        \hH_{\mathbf{q}ac} &= -2 E_J\left[\cos \left(\bphq\right)-1\right] \times \\
        \Big(\cos&\Big\{\bm{\varphi}_a\big[ u_{aa} (\ha+\ha^\dagger + 2 \alpha_a )+ u_{ac} (\hc + \hc^\dagger + 2 \alpha_c )  \big]\Big\}-1\Big).
    \end{split}
\end{align}

We now consider the remainder of the \cref{eq:HaHc}, and perform a time-dependent displacement transformation (\cite{Petrescu2020} App.~B) to eliminate the linear drive term in the remaining linear Hamiltonian
\begin{align}
    \begin{split}
        \hH_a + \hH_c =&  \wa \ha^\dagger \ha+ \wc \hc^{\dagger} \hc \\
        &+ \eps_d (t) (-i)\left[v_{ca}(\ha-\ha^\dagger)+ v_{cc}(\hc-\hc^\dagger)\right].
    \end{split}
\end{align}
The effect of this unitary is to remove the charge drives and to  displace the phase operators in all Josephson terms that is $\ha + \ha^\dagger \to \ha + \ha^\dagger + \eta_{a,\varphi} e^{-i\wdr t} + \eta_{a,\varphi}^* e^{i \wdr t}$ and an analogous equation for $c \leftrightarrow a$, where
\begin{align}
    \begin{split}
    \eta_{j,\varphi} = \frac{v_{cj} \epsilon_{d}\left(\wdr+i \kappa_j/2\right)}{\omega_j^2-\left(\wdr + i \kappa_j/2\right)^2}, \quad j=a,c.
    \end{split}
\end{align}
We have allowed for finite relaxation rates of the ancilla and cavity polaritons. Letting moreover $\eta_{j,n}=-i\eta_{j, \varphi} \omega_j /\left(\wdr+i \kappa_j/2\right) $, we define steady-state photon numbers $\bar{n}_j=\left|\left(\eta_{j, \varphi}+i \eta_{j, n}\right) / 2 \right|^2$. In the limit of high quality factor $\kappa_p \ll \omega_p, \wdr$, we have $\omega_p /\left(\wdr+i \kappa_j/2\right) \approx \omega_j/\omega_d$. Further taking a nearly resonant drive at the frequency of the cavity polariton, we have $|\eta_{a,\varphi}| \approx 0$ and $\eta_{c,\varphi} \approx \sqrt{\bar{n}}$. 

Putting everything together, we  arrive at
\begin{align}
    \begin{split} \label{eq:HsumAgain}
    \hH &= \hH_\mathbf{q} + \hH_a + \hH_c + \hH_{\mathbf{q}ac} + \hH_{ac},
    \end{split}
\end{align}
with the decoupled transmon, ancilla , and cavity polariton modes described by  $\hH_\mathbf{q} = 4 \ECq(\bnq - n_g)^2-2 E_J \cos \left(\bphq\right)$ as defined before, and $\hH_a + \hH_c = \hbar \wa \ha^\dagger \ha + \hbar\wc \hc^\dagger \hc$, and now the interaction Hamiltonians contains flux- and drive-dependent contributions 

\begin{align}
    \begin{split}\label{eq:HintAgain}
        \hH_{ac} &=- 2 E_J \cos\Big\{\bm{\varphi}_a\big[ u_{aa} (\ha+\ha^\dagger) +  u_{ac} (\hc + \hc^\dagger )   \big] \\ &\hspace{60pt}+ \bar{\varphi}_\text{ext} + \tilde{\eta}(t) \Big\},\\
        \hH_{\mathbf{q}ac} &= -2 E_J\left[\cos \left(\bphq\right)-1\right] \times \\ &\;\;\;\;\Big(\cos\Big\{\bm{\varphi}_a\big[ u_{aa} (\ha+\ha^\dagger )+ u_{ac} (\hc + \hc^\dagger  )  \big] \\&\hspace{35pt}+ \bar{\varphi}_\text{ext} + \tilde{\eta}(t) \Big\}-1\Big),
    \end{split}
\end{align}
where the flux dependence enters via $\bar{\varphi}_\text{ext}=2\bm{\varphi}_a (u_{aa} \alpha_a + u_{ac} \alpha_c)$ and the drive contribution is $\tilde{\eta}(t) =  2 \bm{\varphi}_a u_{ac} \sqrt{\bar{n}} \cos(\wdr t)$.

This form, in which several off-diagonal contributions appearing in the linear part of the Hamiltonian have been removed, is well suited for numerical diagonalization. In the subsequent sections, when performing exact diagonalization, we first find the energy eigenstates of $\hH_{\mathbf{q}ac}$. For $\hH_\mathbf{q}$ we find a desired number of lowest $D \gg 2$ energy eigenstates by expressing this decoupled Hamiltonian in the charge basis with $501$ charge operator eigenstates. Then we re-express the interaction Hamiltonian $\hat{H}_{\mathbf{q}ac} + \hat{H}_{ac}$ in this basis (see e.g. \cite{Shillito2022}) and obtain a coupled Hamiltonian with low Hilbert space dimension that reproduces accurately the low-energy states.

The Hamiltonian can be further simplified by freezing the passive ancilla mode. This requires performing a normal-ordered expansion of the part of the Josephson potential involving the ancilla in \cref{eq:HintAgain}, and retaining only the $0^\textit{th}$ order contribution assuming that the ancilla remains in its vacuum state. The resulting model is
\begin{align}
    \label{eq:H_BA}
    \hH = \hH_\mathbf{q} + \hH_c + \hH_{\mathbf{q}c},
\end{align}
with $\hH_\mathbf{q}$ as before,
\begin{align}
    \begin{split}
    \label{eq:HC_BA}
    \hH_{c} =& \hbar \omega_c \hc^\dagger \hc \\
    &- 2 E_J e^{-\bm{\varphi}_a^2 u_{aa}^2/2} \cos \Big[\bm{\varphi}_a u_{ac} (\hc+\hc^\dagger) + \bar{\varphi}_\text{ext} + \tilde{\eta}(t) \Big], 
    \end{split}
\end{align}
and 
\begin{align}
    \begin{split}
    \label{eq:HQC_BA}
        \hH_{qc} =& -2 E_J e^{-\bm{\varphi}_a^2 u_{aa}^2/2}  \left[\cos \left(\bphq\right)-1\right] \times \\&\Big\{\cos \Big[\bm{\varphi}_a  u_{ac} (\hc + \hc^\dagger  ) + \bar{\varphi}_\text{ext} + \tilde{\eta}(t) \Big]-1\Big\}.
    \end{split}
\end{align}
The elimination of the ancilla is advantageous as it allows us to increase the local Hilbert space of the $c-$polariton to allow the branch analysis \cite{Shillito2022} to access the photon numbers used in the experiment. Replacing $\varphi_c = \bm{\varphi}_a u_{ac}$ and defining $\bar{E}_J = e^{-\bm{\varphi}_a^2\frac{u_{aa}^2+u_{ac}^2}{2}} E_J$, we arrive at \cref{eq:2mode_ham} in the main text.

\section{Normal-mode transformation}
\label{Ap:NM}
In this appendix we derive the polaritons as normal-modes of the ancilla-cavity coupled Hamiltonian. To this end, we consider the bare cavity linearly coupled to the linearized ancilla. With subscript `2' for `quadratic', we need to find the normal modes
\begin{align}
    \begin{split}
        H_2/\hbar = \Wa \hba^\dagger \hba + \Wc \hbc^\dagger \hbc - g_{ac} (\hba-\hba^\dagger)(\hbc-\hbc^\dagger),
    \end{split}
\end{align}
where for the bare ancilla mode we only consider the dominant contribution coming from the inductor, i.e. $\Wa = \sqrt{8 E_{La} E_{Ca}}$, with $E_{La} = 2 E_J \frac{L_J}{L_a}$.

Following \cite{Malekakhlagh2020}, App. A.2. the bare mode quadratures write in terms of normal modes as 
\begin{align}
    \begin{split}\label{eq:bn}
        \hba + \hba^\dagger &= \uaa (\ha+\ha^\dagger) + \uac (\hc+\hc^\dagger), \\
        \hbc + \hbc^\dagger &= \uca (\ha+\ha^\dagger) + \ucc (\hc+\hc^\dagger), \\
        (-i)(\hba - \hba^\dagger) &= v_{aa} (-i)(\ha-\ha^\dagger) + v_{ac} (-i)(\hc-\hc^\dagger), \\
        (-i)(\hbc - \hbc^\dagger) &= v_{ca} (-i)(\ha-\ha^\dagger) + v_{cc} (-i)(\hc-\hc^\dagger),
    \end{split}
\end{align}
in terms of eight coefficients to be defined below. Letting 
\begin{align}
    \begin{split} \label{eq:ss}
    \tan (2 \theta)&\equiv\frac{4 g_{ac} \sqrt{\Wa \Wc}}{\Wc^2-\Wa^2}, \\
    \wa &\equiv [ \Wa^2 \cos ^2(\theta)+\Wc^2 \sin ^2(\theta)-2 g_{ac} \sqrt{\Wa \Wc} \sin (2 \theta) ]^{1/2} \\
    \wc &\equiv [ \Wc^2 \cos ^2(\theta)+\Wa^2 \sin ^2(\theta)+2 g_{ac} \sqrt{\Wa \Wc} \sin (2 \theta) ]^{1/2} \\
    s_1 &\equiv \left(\Wc / \Wa \right)^{1 / 4}, 
    s_2 \equiv \left[\frac{\wa^2}{\Wa \Wc}\right]^{1 / 4}, 
    s_3 \equiv \left[\frac{\wc^2}{\Wa \Wc}\right]^{1 / 4},
    \end{split}
\end{align}
the normal-mode coefficients are
\begin{align}
    \begin{split}
        \left[\begin{array}{ll}
            u_{aa} & u_{ac} \\
            u_{ca} & u_{cc}
            \end{array}\right] &= \left[\begin{array}{cc}
                s_1 s_2 \cos (\theta) & s_1 s_3 \sin (\theta) \\
                -s_1^{-1} s_2 \sin (\theta) & s_1^{-1} s_3 \cos (\theta)
                \end{array}\right], \\
        \left[\begin{array}{ll}
                v_{aa} & v_{ac} \\
                v_{ca} & v_{cc}
            \end{array}\right] &= 
            \left[\begin{array}{cc}
                    s_1^{-1} s_2^{-1} \cos (\theta) & s_1^{-1} s_3^{-1} \sin (\theta) \\
                    -s_1 s_2^{-1} \sin (\theta) & s_1 s_3^{-1} \cos (\theta)
                \end{array}
                \right].
    \end{split}
\end{align}
With these transformations, the quadratic Hamiltonian becomes simply, in terms of the normal-mode frequencies, the second and third lines of \cref{eq:ss},
\begin{align}
    H_2/\hbar = \wa \ha^\dagger \ha + \wc \hc^\dagger \hc.
\end{align}

\section{Fabrication process}
\label{apx:samplefab}


The fabrication of the transmon molecules sample starts from a high-resistivity \ce{Si}$\left(100\right)$ two-inch wafer. All structures are patterned in a single step of electron-beam lithography. Beforehand, $8\times\SI{8}{\micro\meter^2}$ square golden markers are patterned on the wafer for alignment and focus of the beam. The wafer is then cleaned in a BOE (Buffer Oxide Etchant) 1:7 solution for \SI{20}{\second} in order to remove any organic impurities as well as the native silicon oxide layer. A double layer of electronic resist is spincoated on the wafer. The first layer is \SI{750}{\nano\meter} thick, made from PMMA/MAA 8\% (ARP 617.14/600.07). The second is a \SI{270}{\nano\meter} layer of PMMA 950K 4\% (ARP.679.04).

The structures are then patterned by a \SI{80}{\kilo\electronvolt} electron-beam lithographer (NanoBeam Ltd.). The inductor $L_a$ is patterned as a chain of SQUIDs made from bridge-free technique \cite{Lecocq2011}. The individual Josephson junctions, of energy $E_J$, are made from Dolan bridge technique \cite{Dolan1977}.

The patterned mask is developed with cold development. We use a 3:1 solution of IPA/$\ce{H}_2\ce{O}$ as developer and cool it to $1.0\pm\SI{0.1}{\celsius}$. The wafer is dipped in the cold developer for \SI{60}{\second}, then rinsed in cold water for at least \SI{30}{\second} and dried. After the development, we perform a \SI{15}{\second} RIE (Reactive Ion Etching) descum at \SI{10}{\watt}.

The structures are evaporated in a single step of double-angle aluminum evaporation. The sample is loaded in an electron-beam evaporation system (PLASSYS) and is not unloaded between the two aluminum evaporations. A vacuum around $\SI{2.9e-8}{\milli\bar}$ is achieved in the chamber and loadlock, by pumping overnight and then performing a titanium flash. The first aluminum layer is \SI{20}{\nano\meter} thick and evaporated at \SI{0.1}{\nano\meter\per\second}, at an angle of \SI{-35}{\degree}. The aluminum is then partially oxided by filling the chamber with \SI{5}{\milli\bar} of $\ce{O}_2$ for 5 minutes. After the oxidation, the chamber is pumped and a titanium flash is performed, leading to a vacuum around $\SI{2.8e-8}{\milli\bar}$. The second layer of aluminum is then evaporated: \SI{50}{\nano\meter} of aluminum are deposited at \SI{0.1}{\nano\meter\per\second}, at an angle of \SI{+35}{\degree}. For capping, a second oxidation is performed, by filling the chamber with \SI{5}{\milli\bar} of $\ce{O}_2$ for 5 minutes.

Lift off is then performed by leaving the wafer in a beaker of NMP heated \SI{80}{\celsius} for about 10 hours. The wafer is then rinsed with Acetone, Ethanol, IPA and then dried. Finally, the 2-inch wafer is diced into chips of dimensions $5\times\SI{6.8}{\milli\meter}$, with a Disco$^{TM}$ DAD 321 Automatic Dicer.

\section{Sample parameters}
\label{apx:sampleparams}

\begin{table}
\centering
\begin{tabular}{c@{\hspace{2em}}c} 
\toprule
\toprule
Parameter name & Value \\
\toprule
$\Wq/2\pi$ & \SI{2.0687}{\giga\hertz} \\
$\wc/2\pi$ & \SI{7.294}{\giga\hertz} \\
$\wa/2\pi$ & \SI{6.502}{\giga\hertz} \\
$\balq/2\pi$ & \SI{-81.4}{\mega\hertz} \\ 
$\chi_{qc}/2\pi$ & \SI{-2.02}{\mega\hertz} \\
$\kappa_c/2\pi$ & \SI{17.2}{\mega\hertz} \\

\midrule
$\ECq/h$ & \SI{0.0734}{\giga\hertz} \\
$\ECa/h$ & \SI{0.0335}{\giga\hertz} \\
$E_J/h$ & \SI{3.96}{\giga\hertz} \\
$L_a$ & \SI{4.24}{\nano\henry} \\
$\Wc/2\pi$ & \SI{7.23}{\giga\hertz} \\
$g_{ac}/2\pi$ & \SI{215}{\mega\hertz} \\

\midrule

$\Wa/2\pi$ & \SI{6.59}{\giga\hertz} \\
$\bala/2\pi$ & \SI{-1.64}{\mega\hertz} \\
$\alc/2\pi$ & \SI{-0.0143}{\mega\hertz} \\
$\ala/2\pi$ & \SI{-1.20}{\mega\hertz} \\
$\chi_{qa}/2\pi$ & \SI{-20.6}{\mega\hertz} \\
$\theta$ & \SI{0.298}{rad} \\

\midrule
$T_1$ & \SI{124.5}{\micro\second} \\
$T_2^*$ & \SI{10.6}{\micro\second} \\
$T_2^E$ & \SI{22.6}{\micro\second} \\

\bottomrule
\bottomrule
\end{tabular}
\caption{\textit{Main parameters of the measured sample and cavity}. The first group, from $\Wq$ to $\kappa_c$, consists of parameters which were measured directly. The second group, from $C_s$ to $g_{ac}$, is devoted to parameters fitted using the model \cref{eq:Htot}, as discussed in \cref{apx:fit}. In the third group, from $\Wa$ to $\theta$, the parameters were derived from the previous quantities. The last group contains measured coherence times. }
\label{tab:target_params}
\end{table}


The main parameters of the measured sample are shown in \cref{tab:target_params}. The frequencies $\Wq$, $\wc$, $\wa$ and the anharmonicity $\balq$ are measured precisely with spectroscopies, at zero flux. The dispersive shift $\chi_{qc}$ and readout mode decay rate $\kappa_c$ are fitted from Ramsey-Stark measurements as a function of readout frequency \cite{Ficheux2018}. The parameters $\ECq$, $\ECa$, $E_J$, $L_a$, $\Wc$ and $g_{ac}$ are fitted simultaneously from measured spectroscopies as a function of the flux, as explained in \cref{apx:fit}. We use the measured and fitted values to derive the remaining parameters $\Wa=(\wc^2+\wa^2-\Wc^2)$, $\theta=(1/2)\arctan[4g_{ac}\sqrt{\Wc\Wa}/(\Wc^2-\Wa^2)]$, $L_J=\varphi_0^2/E_J$, $\bala=-E_{Ca}/[\hbar(1+2L_J/L_a)]$, $\alc=\sin^4\theta(\wc/\Wa)^2\bala$, $\ala=\cos^4\theta(\wa/\Wa)^2\bala$, $\chi_{qa}=\chi_{qc}\wa\cos^2\theta/(\wc\sin^2\theta)$, with the reduced flux quantum $\varphi_0=\hbar/(2e)$.

The coherence times $T_1$, $T_2^*$, $T_2^E$ are characterized through standard relaxation, Ramsey and Hahn-Echo measurements, respectively.

\section{Fitting the spectroscopic data}
\label{apx:fit}
\begin{figure*}
    \includegraphics[width=0.75\linewidth]{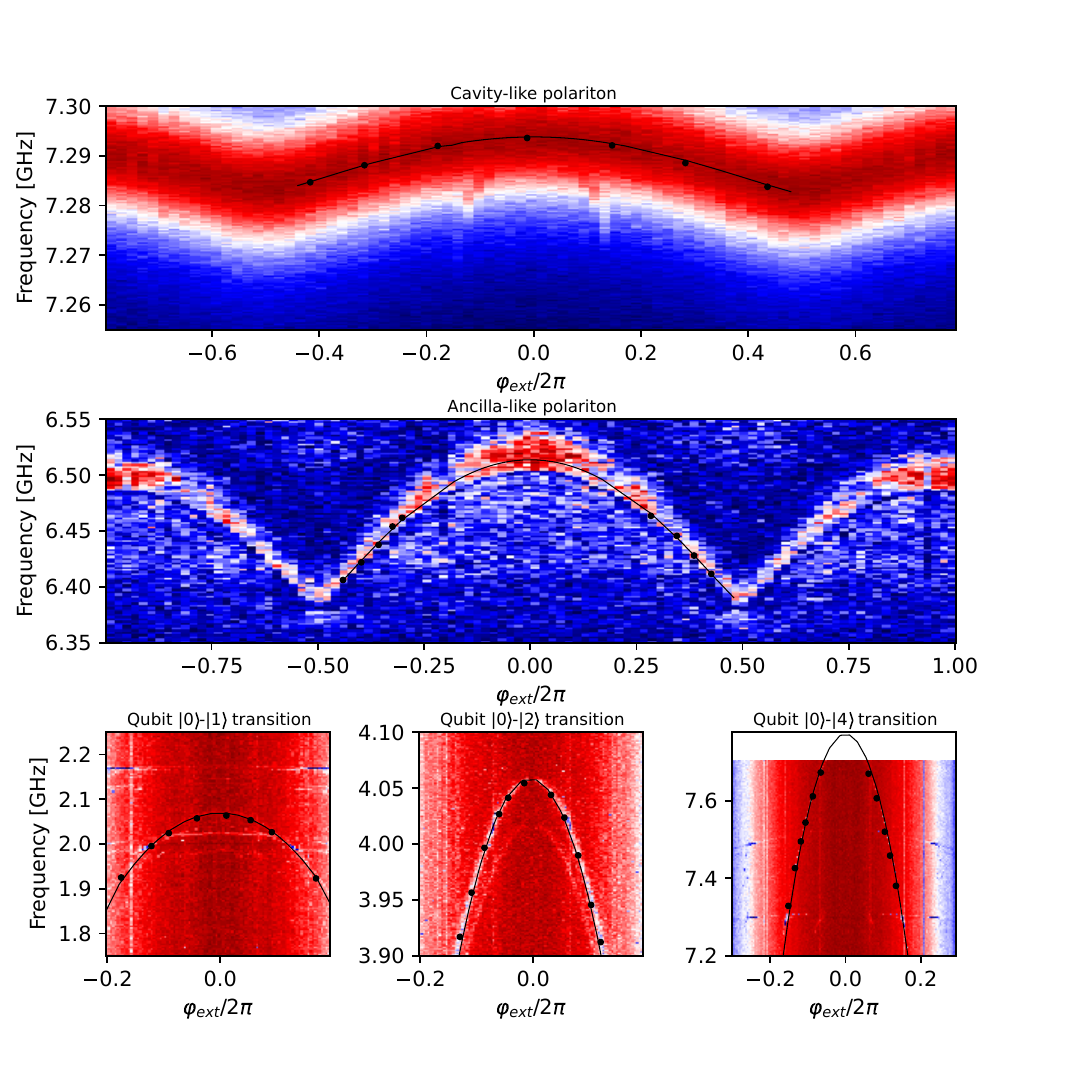}
    \caption{Spectroscopic data (absolute value of the transmission coefficient $|S_{21}|$) for single-excitation manifold of the upper and lower polariton and the transmon qubit mode, along with the transmon 02 and 04 transitions, along with digitized points (black circles) for the fit, and the fit result (solid black curve). 
    \label{fig:Spec}}
\end{figure*}

We fit the spectroscopic data \cref{fig:Spec} using the model of \cref{eq:Htot} at vanishing drive $\bar{n}=0$. To this end, we consider a Hilbert space size of $D=8$ transmon energy eigenstates, and 3 Fock eigenstates for each of the polaritons. We label eigenstates $\ket{\overline{j,n_a,n_c}}$ according the maximized overlap against the product states of the decoupled system $\ket{j}\ket{n_a}\ket{n_c}$. This allows us to define from the model transition frequency functions as a function of a swept parameter, here the flux $\varphi_\textit{ext}$.

We digitize the spectroscopic data using an interactive tool. We optimize for the square of the distance between the corresponding transition frequency obtained from the numerical model and the digitized points, while ensuring that each digitized point is weighed equally. We do this by rescaling the distance between model and each digitized point by the bandwidth of the digitized points corresponding to each of the transitions considered. In the fitting procedure, we consider the single-excitation transition 01 of the cavity and ancilla polaritons, and of the transmon mode, as well as the transmon mode 02 and 04 transitions (see \cref{fig:Spec}). Additionally, we add to the fit cost function the qubit $01$ transition frequency measured from a Ramsey experiment $\Wq^\textit{exp}/2\pi = 2.0687 \text{ GHz }$ with the distance to the numerical model weighted by $5 \text{ MHz }$, the cavity polariton frequency $\omega_{c}^\textit{exp}/2\pi = 7.294 \text{ GHz }$ with the distance to the numerical model weighted by $10 \text{ MHz }$, and the cross-Kerr between the transmon mode and the cavity polariton $\chi_{qc}/2\pi = -2.02 \text{ MHz}$ with a $200 \text{ kHz}$ window. 

With these constraints, using the SLSQP method in \path{scipy.optimize}, we arrive at the model parameters enumerated in \cref{tab:main_params_only,tab:target_params}. The obtained model yields a transmon qubit mode transition frequency $\Wq^\textit{model}/2\pi = \SI{2.0693}{\giga\hertz}$, a transmon 01 transition to cavity polariton cross-Kerr $\chi^\textit{model}_{qc}/2\pi = \SI{-2.032}{\mega\hertz}$, a cavity polariton frequency $\wc^\textit{model}/2\pi = \SI{7.2938}{\giga\hertz}$.

\section{Microwave setup}
\label{apx:setup}

\begin{figure}
\centering
\includegraphics[width=1\columnwidth]{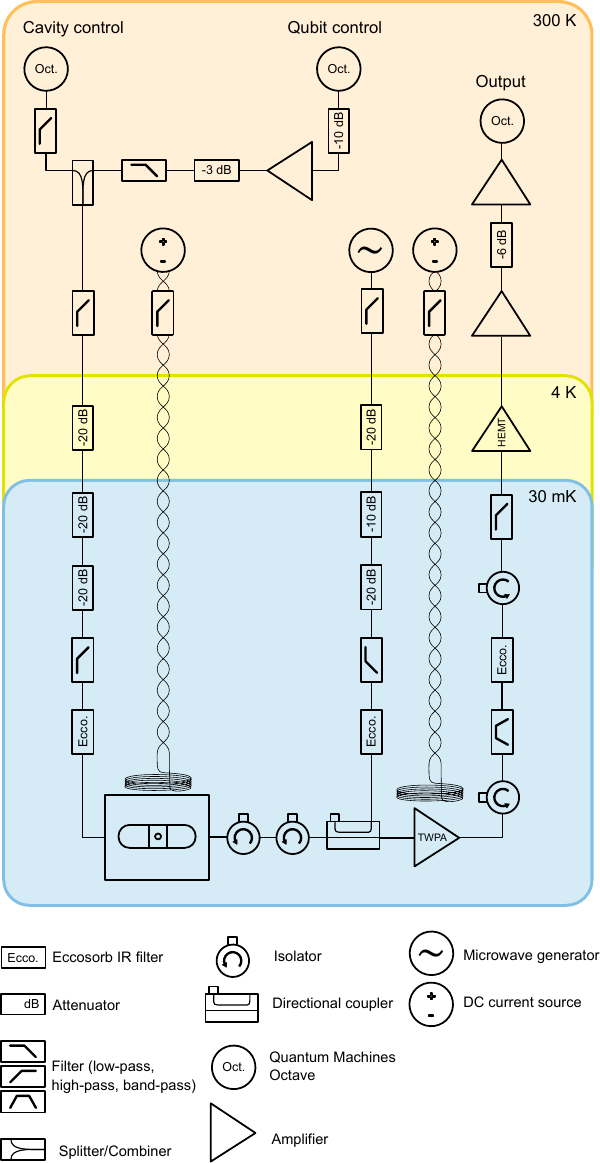}
\caption{\label{apxfig:setup} Full microwave setup. }
\end{figure}


The transmon molecule sample is measured in a homemade dilution refrigerator with a base temperature of \SI{30}{\milli\kelvin}. The full microwave setup is shown in Fig. \ref{apxfig:setup}. The signal generation and acquisition are performed using a Quantum Machines OPX+ at room temperature. A Quantum Machines OCTAVES is responsible for up-conversion and down-conversion of the signals.

A single input line goes into the cryostat, used both for qubit control and for readout. The input signals travel down this input line, undergo several stages of attenuation and filtering, then reach the 3D cavity containing the transmon molecule circuit. The readout signal then leaves the cavity through its output port. On the output line, the signal undergoes different stages of amplification, the first of which is a homemade TWPA with ``reverse-Kerr'' phase matching \cite{Ranadive2022}. It is flux-biased through a superconducting coil connected to a Keysight B2902A current source. The TWPA pump tone is generated by an HP83630A microwave source, and coupled to the output line through a directional coupler. Following the TWPA amplification stage, the signal is filtered and reaches an LNF-LNC1\_12A HEMT at \SI{4}{K}. 

The 3D bulk cavity containing the transmon molecule chip is made of OFHC (Oxygen-Free High thermal Conductivity) copper. The cavity has dimensions $35\times24.5\times\SI{5}{\milli\meter}$ and we use its $\text{TE}_{101}$ mode. A coil of superconducting wire is wound around the cavity in order to flux bias the sample. The bias is applied by a Keysight B2902A current source. The sample is affixed to the cavity and thermalized using indium wire. The cavity's input and output couplings are measured at room temperature as $\kappa_\text{in}/2\pi = \SI{0.153}{\mega\hertz}$ and $\kappa_\text{out} /2\pi = \SI{13.0}{\mega\hertz}$, respectively. The copper cavity is placed inside a half-cylindrical mu-metal shield, with IR-absorbing coating on the inside.

\section{Multi-state single-shot measurements}
\label{apx:iq_plane}

\begin{figure}
\centering
\includegraphics[width=0.9\columnwidth]{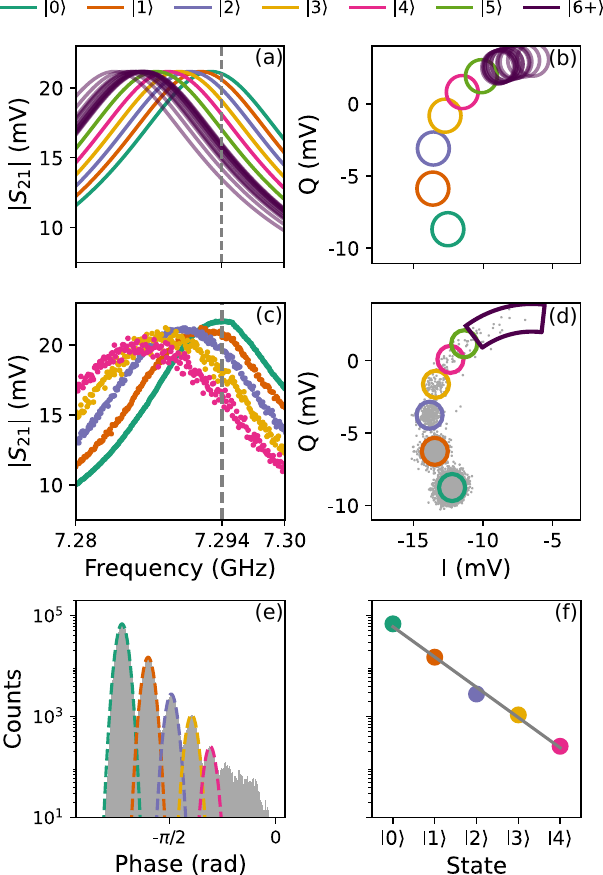}
\caption{\label{apxfig:msss} \textit{Multi-state single-shot readout, example of thermal state measurement}. (a) Simulation of the expected readout mode spectroscopy for different transmon states. Rescaled for easier comparison to the experimental data. (b) Simulation of the expected pointer state positions in the IQ plane using the expected dispersive shift for each transmon state. The circle radii are computed using the experimental SNR. Rescaled for easier comparison to the experimental data. (c) Spectroscopy of the readout mode for a given transmon state. The state preparation is done using preselection. (d) Experimental data for a thermal state measurement plotted in phase space. The overlaid colored lines are the thresholds used to classify the experimental points as different transmon states. (e) Histogram of the thermal state measurement. The IQ plane coordinates are converted to an arbitrary phase varying along the circle described by the centers of the pointer states. (f) Population distribution of the thermal state. Following a thermal distribution of effective temperature \SI{72}{\milli\kelvin}. }
\end{figure}

In this Appendix, we give details on the multi-state single-shot readout technique, which was developed in order to identify and separate as many transmon states as possible in the IQ plane. The number of discernible states is maximized by working in the limit $\chi_{qc}\ll\kappa_c$. In this regime, the dispersive shift induced by transmon state $\ket{k}$ on the readout mode remains smaller than the readout mode width $\kappa_c$, even for $k\gg1$. This form of broadband multi-state readout can be simulated in scQubits \cite{Groszkowski2021}, with the two-mode Hamiltonian of \cref{eq:2mode_ham} and the parameters from \cref{tab:target_params}. The readout mode's transmission profile is modeled as a lorentzian of width $\kappa_c$, as shown in \cref{apxfig:msss}(a). The readout drive frequency is set to $\omega_d=\SI{7.294}{\giga\hertz}$. The magnitude and phase of the transmission at $\omega_d$ give us the coordinates of the pointer states in the IQ plane, shown in \cref{apxfig:msss}(b), where the blob widths were plotted using our experimental SNR. The simulation shows that the pointer states associated to states $\ket{0}$ to $\ket{5}$ are quite well resolved. The pointer states for the higher transmon states $\ket{6}$ to $\ket{19}$, all shown in dark purple, are bunched together and seem to converge towards a given position of the phase space. This guarantees that there is no ambiguity between the identifiable states $\ket{0}$ to $\ket{5}$ and the higher unresolved states.


The simulations from \cref{apxfig:msss}(a,b) can be verified experimentally. The transmission magnitude of the readout mode with an initialized transmon in states $\ket{0}$ to $\ket{4}$, as shown in \cref{apxfig:msss}(c). Furthermore, the IQ plane simulation from \cref{apxfig:msss}(b) can be compared to a simple single shot measurement of the transmon's thermal state, shown in \cref{apxfig:msss}(d). The data shows several pointer states positioned along a curve in the IQ plane, just as expected from the scQubits simulation. We can resolve and fit six of the blobs, corresponding to states $\ket{0}$ to $\ket{5}$, then observe a non-gaussian zone akin to a continuum of points, corresponding to states $\ket{6}$ and higher. In order to extract the population of each state, the separable pointer states $\ket{0}$ to $\ket{5}$ are fitted by 2D gaussians. Circular thresholds are then set for each state, with a radius $2\sigma$, where $\sigma$ is the width of the gaussian, as seen in \cref{apxfig:msss}(d). We also add a separate much wider threshold for all the higher non-resolved states, noted $\ket{6+}$. The points which are categorized nowhere are classified as outliers and correspond to points which are difficult to categorize due to SNR limitations.

As an example of application, the multi-state single-shot readout can be used to precisely fit the transmon's effective temperature. The points from the measured thermal state distribution, in \cref{apxfig:msss}(d), are projected along the curve defined by the centers of the pointer states and represented as a 1D histogram, seen in Fig. \ref{apxfig:msss}(e). The first five pointer states are fitted with gaussians in order to get their populations. The populations are then plotted as a function of the associated transmon state, see \cref{apxfig:msss}(f). These populations are fitted a thermal distribution of effective temperature $T_{\textbf{eff}}=\SI{72}{\milli\kelvin}$. Note that the population of state $\ket{5}$ was not fitted in this case due to its partial overlap with higher states, which could bias the fit of the effective temperature.

In the case of MIST measurements, the multi-state single-shot readout thresholds are used to extract the probability to be in each identifiable transmon state or in $\ket{6+}$. For computing the probabilities, we take out the outliers from the statistic and renormalize by the remaining total population.

\section{CLEAR pulse-shaping}
\label{apx:clear}

\begin{figure}
\centering
\includegraphics[width=0.9\columnwidth]{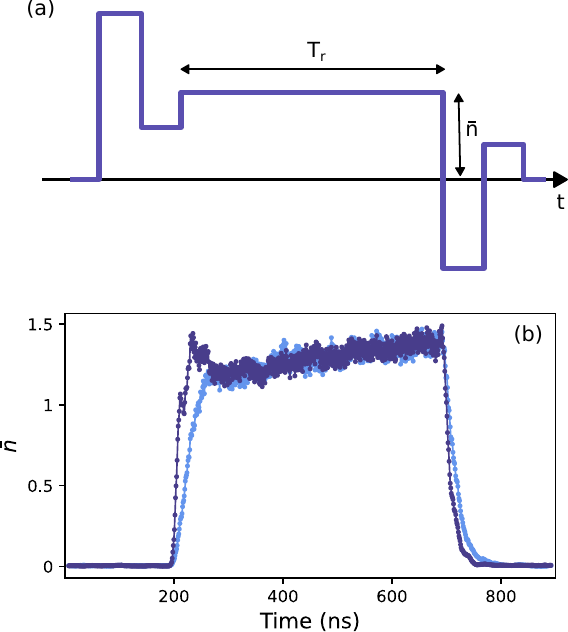}
\caption{\label{apxfig:clear} \textit{CLEAR pulse-shaping}. 
(a) Scheme of the CLEAR pulse shape. (b) Demodulated output signal's envelope when sending a square (in light blue) or a CLEAR (in dark blue) readout pulse. The pulse duration is $T_r=\SI{500}{\nano\second}$. }
\end{figure}

In our MIST measurements, the cavity drive pulse is shaped as a CLEAR pulse \cite{McClure2016}. The detailed shape of the pulse is shown in \cref{apxfig:clear}(a). The ring-up is overshot for a short duration, then over-compensated, before reaching its target power $\bar{n}$. The ring-down also follows the same procedure before stabilizing at zero power. This pulse shape results in a faster ring-up and ring-down of the microwave pulse compared to a standard square pulse shape. The overshots' magnitude and durations are optimized for our readout mode's decay rate $\kappa_c$. The comparison between square and CLEAR pulse shape is verified experimentally by monitoring the signal output of the cavity when a cavity drive pulse is sent, as seen in \cref{apxfig:clear}(b). The square pulse presents a ring-up time of \SI{26}{\nano\second} and the ring-down of \SI{34}{\nano\second}. In contrast, the CLEAR pulse presents a ring-up of only \SI{4.6}{\nano\second} and a ring-down of \SI{25}{\nano\second}. Note that the asymmetry between ring-up and ring-down has also been measured in Ref. \cite{Hazra2024} and is believed to stem from the high power of the drive.

\section{Number of photon calibration}
\label{apx:nphcalib}


\begin{figure}
\centering
\includegraphics[width=0.9\columnwidth]{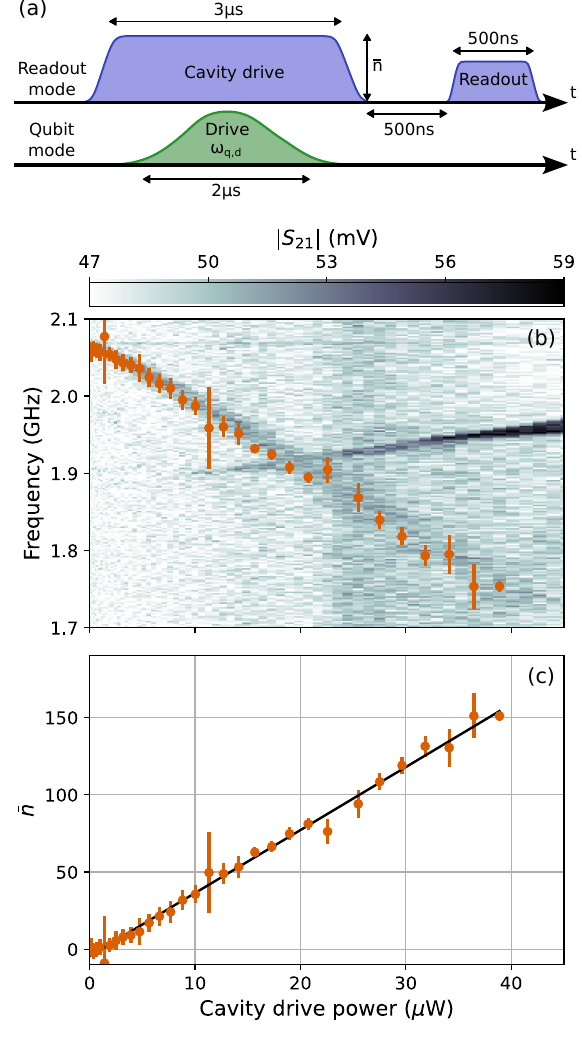}
\caption{\label{apxfig:nph_calib} \textit{Number of photon calibration}. 
(a) Pulse sequence for the AC Stark shift measurement. (b) Spectroscopy of the transmon's frequency $\Wq$ as a function of the power of the cavity's drive. The transmon's frequency is fitted by a lorentzian for each $\bar{n}$, with the fitted centers shown as orange dots, along with the fits' errorbars. (c) Linear calibration of the number of photons in the readout mode $\bar{n}$ as a function of the power of the cavity drive. }
\end{figure}


This Appendix details the calibration procedure used to convert the cavity drive power $P$ into an number of photons $\bar{n}$. To be more exact, $\bar{n}$ represents the number of photons in the readout mode at frequency $\omega_d=\SI{7.294}{\giga\hertz}$, when the transmon is in state $\ket{0}$, which is the same convention as most works in the state-of-the-art \cite{Walter2017,Swiadek2023a}. 

The calibration is performed through a measurement of the AC Stark shift of the transmon frequency $\Wq$, as was done in \cite{Schuster2005,Sank2016}. The pulse sequence, shown in \cref{apxfig:nph_calib}(a), consists in performing a pulsed qubit spectroscopy in the presence of a readout drive at frequency $\omega_d=\SI{7.294}{\giga\hertz}$. The frequency of the qubit drive and the power of the cavity drive are swept, giving the final result shown in \cref{apxfig:nph_calib}(b). The transmon frequency visibly shifts linearly as a function of the power of the cavity drive. Around \SI{1.95}{\giga\hertz}, where we observe an anti-crossing that could be explained by a coupling to a parasitic TLS. At high powers, the transition reaches \SI{1.7}{\giga\hertz} and becomes too low-frequency to be driven, since the qubit drive is sent through the 3D cavity's input port. The resonance $\Wq$ is fitted with a lorentzian for each power value, giving a calibration of the transmon frequency AC Stark shift $\Delta\Wq(P)=\Wq(P)-\Wq(0)$ as a function of the cavity drive power $P$.

The AC Stark shift is converted to an average number of photons through the simple linear relation $\bar{n}(P)=\Delta\Wq(P)/\chi_{qc}$. This requires a precise measurement of the dispersive shift $\chi_{qc}$, which is fitted through Ramsey-Stark measurements up to about $1\%$ uncertainty. The resulting calibration of the number of photons as a function of cavity drive power is shown in \cref{apxfig:nph_calib}(c) and shows linear behavior in all the measured power range, up to $150$ readout photons. This calibration is linearly extrapolated to higher powers for the scales of the MIST measurements.

\section{Studies of structural stability}
\label{Ap:Stability}

In this section, we consider the problem of structural stability of the transmon molecule circuit. We use the simplified model of \cref{eq:H_BA}, \cref{eq:HC_BA}, \cref{eq:HQC_BA}. This lighter two-mode model allows exact diagonalization to large cavity polariton photon numbers, leading to a branch analysis \cite{Shillito2022}. Then after making a semiclassical approximation on the cavity polariton mode, the second simplification is to study the onset of chaos in the resulting classical $\cos\varphi$-driven transmon molecule. 

For the semi-classical approximation, we freeze the quantum fluctuations of the cavity polariton mode, which yields a transmon Hamiltonian with a Floquet drive:
\begin{align}
    \begin{split} \label{eq:DTM}
        \hH_\mathbf{q}(t) =& 4 \ECq(\bnq - n_g)^2 -2 \bar{E}_J \cos\left[ \tilde{\eta}(t) \right] \cos \left(\bphq\right) .
    \end{split}
\end{align}

\subsection{Branch analysis}
\label{Ap:BA}
For the branch analysis studies presented in the main text, we have checked that Floquet simulations (not presented here) of \cref{eq:DTM} agree qualitatively with the results of \cref{eq:H_BA}, as expected \cite{Cohen2023,Dumas2024}, and with the branch analysis of three-mode Hamiltonian \cref{eq:HsumAgain}. We commit to \cref{eq:H_BA} in order to achieve high photon numbers in the cavity polariton so as to compare to experiment, on the reasonable assumption that the ancilla mode is not populated. We do note that resonances between cavity polariton and ancilla polariton are possible at very high photon numbers, but we found these to not affect the overall branch analysis, and we shall not discuss these further within this work.

Considering thus \cref{eq:H_BA} (same as \cref{eq:2mode_ham} in the main text), we employ the labeling procedure of \cite{Shillito2022} to re-express 
$\hH = \sum_{j,n_c \geq 0} E_{\overline{j,n_c}} \ket{\overline{j,n_c}} \bra{\overline{j,n_c}}$.
 For the exact diagonalization we take local Hilbert space dimensions $D = 20, d_c =500$, amounting to a Hilbert space dimension of $D d_c=10000$. 

\begin{figure}[t!]
    \centering
    \includegraphics[width=\linewidth]{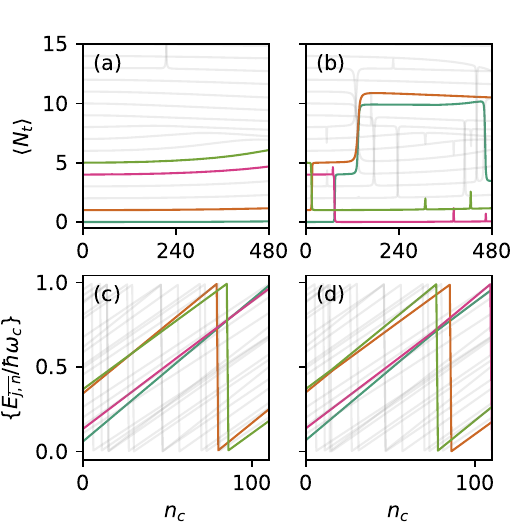}
    \caption{Branch analysis of $\cos\varphi$-coupled circuit at $\varphi/2\pi=0$ (a), and $0.0397$ (b), together with corresponding quasienergy spectra, represented as the fractional part of $E_{\overline{j,n}}/\hbar\omega_c$, in (c) and (d), respectively. Color highlights for $j=0,1,4,5$ following conventions of \cref{fig:mist_meas}(b)-(e). Exact crossings appearing at finite photon number $n_c$ between states transmon states $0$ and $4$, and $1$ and $5$ respectively, at zero flux (c) turn into avoided crossings at nonzero flux (d).}
    \label{fig:branches}
\end{figure}

In \cref{fig:branches} we plot the expectation value of $N_t$ in those states corresponding to transmon states $j=0,\ldots,15$ (with 0, 1, 4, 5 highlighted in color),  versus cavity polariton occupancy $n_c=0,\ldots, 475$. Resonances in the spectrum are marked by places where the values of $N_t$ become equal between different branches [panels (a) and (b)], which is a marker of an avoided crossing in the spectrum [panels (c) and (d)]. These avoided crossings occur between states $\ket{\overline{4,n_c}}$ and $\ket{\overline{0,n_c+1}}$ at a cavity polariton occupancy $n_c$ which is a function of the external flux (similar considerations apply for the crossing $\ket{\overline{5,n_c}}$ and $\ket{\overline{1,n_c+1}}$, which occurs at lower $n_c$ and lower flux). At 0 external flux, this is an exact crossing (see \cref{fig:branches}(c)) leading to no distinguishable feature in the expectation values of $N_t$. Above zero flux, for example at $\varphi_\textit{ext}/2\pi = 0.0397$, the aforementioned exact crossings turn into avoided crossings (d), with marked features in the plots of the expectation value of $N_t$ (the interchange occuring in the pair of curves representing the expectation values of $N_t$ in panel (b) indicates a diabatic traversal of the avoided crossing). As flux is increased, first the transition $\ket{\overline{5,n_c}}$ and $\ket{\overline{1,n_c+1}}$ becomes impossible, as the undriven $\ket{1}-\ket{5}$ transition frequency in the transmon molecule spectrum falls below the cavity polariton frequency. At even larger fluxes the $\ket{0}-\ket{4}$ transition falls below the cavity polariton frequency. This gives rise to the data in \cref{fig:mist_vs_flux}  in the main text.

\subsection{Comparisons to a transversely coupled transmon}
\label{apx:eq_transverse}
In the main text we have compared the $\cos\varphi$-coupled transmon molecule of \cref{eq:H_BA} to a transversely coupled transmon Hamiltonian. Here, for completeness, we provide the details for the analysis of the latter, which are standard \cite{Shillito2022,Cohen2023,Dumas2024}. For the transversely coupled transmon we consider the Hamiltonian
\begin{align}
    \begin{split} \label{eq:Htransverse}
        \hH = \hH_q + \hH_c + \hH_{qc},
    \end{split}
\end{align}
where $\hH_q$ is defined in the first \cref{eq:selfH} (for consistency with the transmon molecule, we keep $2 E_J$ for the Josephson energy), and
\begin{align}
    \begin{split} 
        \hH_c =& \wc \hc^\dagger \hc - i \hbar\varepsilon_d(t) (\hc-\hc^\dagger),\\
        \hH_{qc} =& - i \hbar g_{qc} \bnq (\hc-\hc^\dagger),
    \end{split}
\end{align}
with $\varepsilon_d(t)$ as given before. We begin by perform a displacement transformation, to remove the linear term on the cavity, and transfer it to a charge drive on the transmon, then neglect the quantum fluctuations of the cavity in a semiclassical approximation \cite{Cohen2023} and arrive at the Floquet Hamiltonian for the capacitively driven transmon
\begin{align}
    \begin{split} 
        \hH_q(t) = &4 \ECq(\bnq - n_g)^2 - 2 E_J \cos(\bphq) \\
        &- 2 \hbar g_{qc} \sqrt{\bar{n}} \cos(\wdr t) \bnq.
    \end{split}
\end{align}
We further perform a unitary transformation via $\hat{U}(t) = e^{-i \frac{2 g_{qc} \sqrt{\bar{n}}}{\wdr}  \sin(\wdr t) \bnq } \equiv e^{-i \tilde{\eta}_t(t) \bnq}$, so that 
\begin{align}
    \begin{split} \label{eq:DTT}
    \hH'_q(t) \equiv& \hat{U}(t)\left[ \hH_q(t) - i\partial_t \right] \hat{U}(t)^\dagger \\
    =&4 \ECq(\bnq - n_g)^2 - 2 E_J \cos\left[\bphq - \tilde{\eta}_t(t)\right].
    \end{split}
\end{align}
In order to perform a comparison, we ensure that $g_{qc}$ above is chosen such that the AC Stark shift obtained in the transverse model at a given cavity  photon number matches the one in the transmon molecule model at an equal cavity polariton photon number. To this end, we perform a Jacobi-Anger expansion \cite{Petrescu2023} and solve for $g_{qc}$ imposing that the equivalent Josephson energies coming from the zeroth harmonic contribution in that expansion match between the two models \cref{eq:DTT} and \cref{eq:DTM} at zero flux. Neglecting the dressing of the Josephson energy by the zero-point fluctuations of the polaritons $\bar{E}_J\approx E_J$, 
we require that the renormalization to the Josephson energy from the drive be equal at equal drive photon number in the two models, \cref{eq:DTT} and \cref{eq:DTM}, that is 
\begin{align}
    \begin{split}
        E_J J_0\left(2 g_{qc} \sqrt{\bar{n}}/\wdr\right) =  E_J J_0 \left( 2 \varphi_c \sqrt{\bar{n}}  \right),
    \end{split}
\end{align}
leading to a simple expression for the transverse coupling
\begin{align}
    \begin{split}\label{eq:Match}
        g_{qc} = |\varphi_c| \wdr =  |\varphi_c| \omega_c.
    \end{split}
\end{align}
We have numerically checked that making this choice ensures good agreement in transmon AC Stark shift $E_{\overline{1,n_c}}-E_{\overline{0,n_c}}$ between the transversely coupled model \cref{eq:Htransverse} and the $\cos\varphi$-coupled model with frozen ancilla \cref{eq:H_BA}, as used in the branch analysis plots of \cref{fig:BrPo}. 

\subsection{Classical dynamics}
\label{Ap:Poin}
In this section we consider the stability towards classical chaos of the $\cos\varphi$-coupled or transversely coupled transmon. We perform this analysis on classical limits of the corresponding transmon-mode Hamiltonians: $\cos\varphi$-coupled transmon \cref{eq:DTM}, or transversely coupled transmon \cref{eq:DTT}. In the following, all observables are replaced by classical coordinates without hats. Expanded over harmonics of the drive, the classical Hamiltonians take the same form
\begin{align}
    \mathcal{H}(t) = 4 E_{Cq} \bm{n}_q^2 - \sum_{n=-\infty}^\infty A_n \cos( \bm{\varphi}_q - n \wdr t), \label{eq:GenH}
\end{align}
where the coefficients $A_n$ depend on the type of readout coupling. \Cref{eq:GenH} is a general form for the semi-classical Hamiltonians \cref{eq:semiclass_transverse} and \cref{eq:semiclass_cosphi} in the main text. For the transversely coupled transmon \cref{eq:DTT}
\begin{align}
A_n = 2 E_J J_n(\tilde{\eta}_{t,0}), 
\end{align}
with $\tilde{\eta}_{t,0} = 2g_{qc}\sqrt{\bar{n}} /\omega_d$, whereas for the $\cos-$coupled transmon \cref{eq:DTM} we have
\begin{align}
    \begin{split}
    A_{2n} &= (-1)^n 2 E_J \cos(\bar{\varphi}_\text{ext})  J_{2n}(\tilde{\eta}_0), \\
    A_{2n-1} &= (-1)^n 2 E_J \sin(\bar{\varphi}_\text{ext})  J_{2n-1}(\tilde{\eta}_0),
    \end{split}
\end{align}
where $\tilde{\eta}_0 = 2 \varphi_c \sqrt{\bar{n}}$ is the dimensionless drive amplitude. Note that for the $\cos\varphi$-coupled transmon, odd-integer harmonics only contribute if the flux $\bar{\varphi}_\text{ext} \neq 0$. At zero flux for the $\cos\varphi$-coupled transmon, only the even harmonics are present, as opposed to the transversely coupled transmon. We argued in the main text, by invoking a heuristic criterion for the appearance of chaos summarized below, that the former is more robust against the development of a chaotic layer than the latter.

To characterize the appearance of chaos we recall Chirikov's  heuristic criterion of overlapping resonances \cite{Chirikov1960} used in the main text. According to this, one identifies a change of variables that locally describes the dynamics \cref{eq:GenH} in terms of an undriven pendulum. These regions in phase space are determined by resonances between the  phase oscillation frequency $\dot{\bm{\varphi}}_q$ and a harmonic of the drive $n \wdr$. Chirikov's criterion
states that whenever the separatrices corresponding to different resonances overlap, chaotic dynamics can occur.  The principal resonance is given by the condition $\dot{\bm{\varphi}_q} \approx 0$, which leads to a stationary approximation to \cref{eq:GenH}, $\mathcal{H}_0 = 4 \ECq \bm{n}_q^2 - A_0 \cos(\bm{\varphi}_q)$. Bounded and unbounded trajectories under the Hamiltonian $\mathcal{H}_0$ are separated in phase space by the separatrix curve corresponding to the oscillator being released from the position of highest potential energy with zero kinetic energy $\{ (\bm{\varphi}_q,\bm{n}_q) \mid \mathcal{H}_0 = |A_0| \}$, given by the two curves 
$\bm{n}^{\pm}_0(\bm{\varphi}_q) = \pm \sqrt{|A_0|(1 + \cos\bm{\varphi}_q) /(4 \ECq) }.$ This separatrix has a maximum width along the charge direction of $\Delta \bm{n}_0 =  \sqrt{2|A_0|/\ECq}$ at $\bm{\varphi}_q = 0$. 

Each subsequent harmonic of the drive gives rise to a resonance. Changing coordinates to $\bm{\psi}_m(t) = \bm{\varphi}_q - m \wdr t$, and $\bm{n}_m = \bm{n}_q - m\wdr/(8 \ECq)$, the dynamics close to the $m^{\textit{th}}$ resonance, $\dot{\bm{\psi}}_m \approx 0$, obeys the following stationary approximation of \cref{eq:GenH}
\begin{align}
    \mathcal{H}_m = 4 \ECq \bm{n}_m^2 - A_m \cos(\bm{\psi}_m),
\end{align}
for each integer $m$. Arguments analogous to those given for the principal resonance yield curves for the $m^{\textit{th}}$ separatrix $\bm{n}_m(\bm{\psi}_m) =  \pm \sqrt{|A_m|(1 + \cos\bm{\psi_m})/4 \ECq}.$ Thus the $m^\textit{th}$ separatrix is centered at $m\wdr / (8 \ECq)$ and has a width of $\Delta \bm{n}_m = \sqrt{2 |A_m| / \ECq}$.

\subsection{Constraints on cavity frequency from Chirikov's criterion}
\label{Ap:Constraints}
Chirikov's heuristic criterion  \cite{Chirikov1979May,Lichtenberg} states that a chaotic layer can appear wherever the  separatrices corresponing to stationary resonances approach overlap. This criterion is known to be too crude, in the sense that chaotic regions can appear at lower drive strengths. Thus to preclude chaos, the drive-amplitude-dependent widths of consecutive resonances must be such that the separatrices are far from overlap. For the transmon molecule, this translates to ensuring that the separatrices corresponding to harmonics $0$ and $\pm 2\wdr$ are far, that is
\begin{align}
    \label{eq:cond_chirikov}
    \frac{\omega_d}{\omega_p} \gg 
    \sqrt{|J_0(2\varphi_c \sqrt{\bar{n}})|}+\sqrt{|J_2(2\varphi_c \sqrt{\bar{n}})|} ,
\end{align}
in terms of the equivalent plasma frequency $\omega_p/2\pi \equiv \sqrt{16 E_J \ECq}/h \approx$ \SI{2.156}{\giga\hertz} and $2 \varphi_c \approx 0.0632$ for the fit parameters listed in \cref{apx:fit}. For those parameters, the condition of \cref{eq:cond_chirikov} safely holds for a ratio of between left-hand and right-hand side of lhs/rhs $\geq 2.75$ with $\bar{n} \leq 750$. This is consistent with the absence of chaos in the Poincaré section of \cref{fig:BrPo}(b). For a transversely coupled transmon, however, the equivalent condition is written in terms of the $0^\textit{th}$ and the first odd harmonic $\pm \wdr$, that is 
\begin{align}
    \label{eq:cond_chirikov_transverse}
    \half \frac{\omega_d}{\omega_p} \gg 
    \sqrt{|J_0(2\varphi_c \sqrt{\bar{n}})|}+\sqrt{|J_1(2\varphi_c \sqrt{\bar{n}})|}, 
\end{align}
taking $g_{qc} = |\varphi_c| \wdr$. The condition of \cref{eq:cond_chirikov_transverse} is satisfied only marginally since lhs/rhs $\geq 1.1$ in the same photon-number range $\bar{n} \leq 750$. All of these observations consistent with the appearance of a chaotic layer in \cref{fig:BrPo}(d) at sufficiently large photon number, but not in \cref{fig:BrPo}(b).

\begin{figure}[t!]
\centering
\includegraphics[width=1\columnwidth]{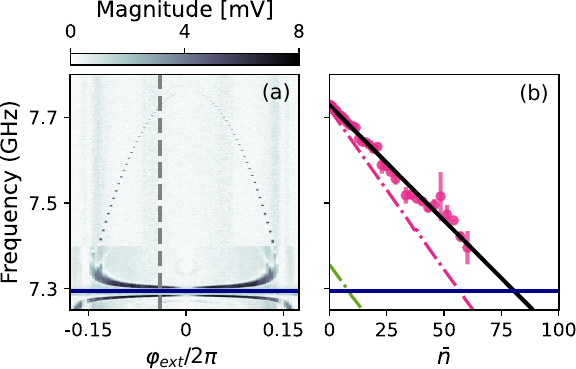}
\caption{\label{apxfig:spectro04} \textit{Spectroscopy and AC Stark shift of the qubit $\ket{0}$ to $\ket{4}$ transition}. 
(a) Spectroscopy of the qubit's $\ket{0}$ to $\ket{4}$ transition as a function of the applied magnetic flux. The vertical grey dotted line indicates the flux bias $\Phi=-0.04\Phi_0$. The dark blue horizontal line displays the readout mode frequency $\wc$. (b) AC-Stark shift measurement of the qubit's $\ket{0}$ to $\ket{4}$ (pink) transition in presence of a cavity drive of variable power $\bar{n}$. The flux bias is fixed at $\Phi=-0.04\Phi_0$. Simulated AC Stark shifts for the $\ket{0}-\ket{4}$ and $\ket{1}-\ket{5}$ transitions in pink and green dash-dotted lines, respectively. }
\end{figure}

\section{Measurement of the \texorpdfstring{$\ket{0}-\ket{4}$}{|0⟩-|4⟩} MIST}
\label{apx:04mist_explan}



In this Appendix, we directly measure the spectral resonance associated to the $\ket{0}-\ket{4}$ MIST. Previous measurements and simulations indicate that the $\ket{0}-\ket{4}$ MIST is an direct transition between transmon states $\ket{0}$ and $\ket{4}$. The frequency of such a transition, noted $\omega_{04}$, can be directly characterized by a two-tone spectroscopy as a function of the applied flux, as shown in \cref{apxfig:spectro04}(a). The measurement reveals that $\omega_{04}$ is quite close to the readout mode frequency $\wc$. For example, at $-0.04\Phi_0$ flux bias, we have $\omega_{04}/2\pi = \SI{7.728}{\giga\hertz}$, resulting in a detuning to $\wc$ of $|\wc - \omega_{04}|/2\pi = \SI{434}{\mega\hertz}$. At fixed flux $-0.04\Phi_0$, we measure the AC Stark shift on $\omega_{04}$ using the same pulse sequence as in \cref{apxfig:nph_calib}(a). The measured AC Stark shift, shown in \cref{apxfig:spectro04}(b), linearly goes to lower frequencies as a function of the photon population up to at least 60 photons. By linearly extrapolating the frequency shift, a crossing between $\omega_{04}$ and $\wc$ is predicted at 81 photons, which is exactly consistent with the power at which the $\ket{0}-\ket{4}$ MIST is measured at flux $-0.04\Phi_0$ in Fig. \ref{fig:mist_meas}(d).

The flux dependence of the $\ket{0}-\ket{4}$ MIST in Fig. \ref{fig:mist_vs_flux}(a) can also be explained from the spectroscopy measurements. As the flux bias gets further from the sweetspot, the transition frequency $\omega_{04}$ is shifted to lower frequencies. This reduces the detuning $|\omega_{04} - \wc|$, meaning that less photons are required to AC-Stark shift $\omega_{04}$ into resonance with $\wc$. At $\pm 0.16$ flux, the frequency $\omega_{04}$ is already in resonance with $\wc$ even with no photons in the readout mode and the $\ket{0}-\ket{4}$ MIST goes to zero photons in \cref{fig:mist_vs_flux}(a). The same type of explanation holds for the $\ket{1}-\ket{5}$, associated to a transition of frequency $\omega_{15}$. At $-0.04\Phi_0$, this frequency is expected to come into resonance with $\wc$ in the presence of 8 photons, as shown by the AC Stark shift simulation in \cref{apxfig:spectro04}(b).

\bibliography{bibliography}

\end{document}